\documentclass[prd,aps,preprint,amsmath,nofootinbib,amssymb,eqsecnum,showkeys,tightenlines]{revtex4-1}
\usepackage{tabularx}
\pdfoutput=1
\usepackage{slashed}
\usepackage{epsfig,latexsym,cancel,amssymb,amsmath,verbatim,mathrsfs}
\usepackage{color}
\usepackage{graphicx}
\usepackage{bm}
\usepackage{tabularx}
\usepackage{amsmath}
\usepackage{amssymb}
\usepackage{amsfonts}
\usepackage{cases}
\usepackage{cancel}
\usepackage{hyperref}
\usepackage{ulem}
\usepackage{here}
\usepackage{color}
\usepackage{graphicx}
\usepackage{diagbox}
\usepackage{multirow}

\def\ra{\rightarrow}
\def\L{\left(}
\def\R{\right)}
\def\wt{\widetilde}

\def\ld{\lambda}
\def\f{\frac}
\newcommand{\be}{\begin{equation}}
\newcommand{\ee}{\end{equation}}
\newcommand{\bea}{\begin{eqnarray}}
\newcommand{\eea}{\end{eqnarray}}
\newcommand{\ba}{\begin{array}}
\newcommand{\ea}{\end{array}}

\long\def\symbolfootnote[#1]#2{\begingroup%
\def\thefootnote{\fnsymbol{footnote}}\footnote[#1]{#2}\endgroup}

 \DeclareMathOperator{\diag}{diag}

\newcommand{\beq}{\begin{equation}}
\newcommand{\eeq}{\end{equation}}
\begin{document}

\title{Massive gauge theory with quasigluon for hot $SU(N)$: Phase transition and thermodynamics}

%as a Unified Picture  Polyakov Background 
\author{Jiang Zhu}
\email[E-mail: ]{jackpotzhujiang@gmail.com}
\affiliation{School of physics, Huazhong University of Science and Technology, Wuhan 430074, China}

\author{Jun Guo}
\email[E-mail: ]{jguo\_hep@163.com}
\affiliation{College of Physics and Communication Electronics, Jiangxi Normal University, Nanchang 330022, China}

\author{Zhaofeng Kang}
\email[E-mail: ]{zhaofengkang@gmail.com}
\affiliation{School of physics, Huazhong University of Science and Technology, Wuhan 430074, China}

\date{\today}

\begin{abstract}

It is challenging to build a model that can correctly and unifiedly account for the deconfinement phase transition and thermodynamics of the hot $SU(N)$ pure Yang-Mills (PYM) system, for any $N$. In this article, we slightly generalize the massive PYM model to the situation  with a quasigluon mass $M_g(T)$ varying with temperature, inspired by the quasigluon model. In such a framework, we can acquire an effective potential for the temporal gauge field background by perturbative calculation, rather than adding by hand. The resulting potential works well to describe the behavior of the hot PYM system for all $N$, via the single parameter $M_g(T)$. Moreover, under the assumption of unified eigenvalue distribution, the $M_g(T)$ fitted by machine learning is found to follow $N$-universality.

%We demonstrate this in a minimal model with  $X$ and the scalon $S$. 

\end{abstract}

\pacs{12.60.Jv,  14.70.Pw,  95.35.+d}

\maketitle

\section{Introduction}

%The Pure Yang-Mills(PYM) gauge theory is the foundation of the particle physics. The high energy behavior of this theory has been widely studied in the past half century and we have successfully established a framework for dealing with it's high energy problems. The PYM sector surviving at low energy is also interest in the new physics domain, in the context of dark sector of the universe, including dark matter (see, e.g., a review \cite{} and also \cite{}) and also in the string theory \cite{}. However, how to deal with this theory at low energy, especially the confinement problem, is still a open question. There are many physicists try to solve this problem: some of them start from the Lagrangian of the PYM field\cite{}, some of them focus on phenomenological potential\cite{} and the holographic theory can also help us to study this problem\cite{}. 

To build a model that can describe the deconfinement phase transition of the $SU(N)$ PYM system at finite temperature, which is hampered by the non-perturbative effect, one should first figure out what knowledge we have about such a system. In the very high-temperature region, it should recover the Stefan-Boltzmann (SB) limit, following the asymptotic freedom of the non-Abelian gauge theory. The more crucial information comes from the lattice simulations, which provide a reliable way to deal with strong coupling, thus furnishing the order of deconfinement phase transition and as well data of therm-dynamic observables, such as pressure and the latent heat $L$~\footnote{The latent heat is not only important to describe the first order phase transition (FOPT) but also critical in cosmology because the Gravitational-wave produced during the deconfinement phase transition in the early universe is directly related to this quantity~\cite{Caprini:2015zlo,Kubo:2018vdw,Helmboldt:2019pan,Halverson:2020xpg,Huang:2020crf,Kang:2021epo}.}. The deconfinement phase transition is a crossover for $N=2$ and FOPT for $N\geq 3$. Moreover, combining the data for pressure and the latent heat allows us to extract the following large $N$ scaling law~\cite{Kang:2021epo}
\begin{equation}
\begin{split}
    p_M=\frac{M^2-1}{N^2-1}p_N,\ \ \ \ L_M=\frac{M^2-1}{N^2-1}L_N,
\end{split}
\end{equation}
where $N$ and $M$ represent different color number. It is challenging to build a model with strong theoretic ground that can correctly account for all of the above aspects of the hot PYM system, for any $N$ beyond $N=3$. However, it is very  meaningful, not only in the theoretical sense but also in the application to the new physics domain, where an (almost) pure $SU(N)$ gauge sector receives wide interest~\cite{baryon,twin,string,Kang:2019izi,Carenza:2022pjd,Sannino:2002wb}. Recently, the prospects of gravitational wave signals during the deconfinement phase transition are studied based on different models~\cite{Kubo:2018vdw,Halverson:2020xpg,Huang:2020crf,Kang:2021epo,Morgante:2022zvc,He:2022amv}. 

%The finial information we should take from lattice result is the order of phase transition. 

The popular line is following the $Z_N$ center symmetry and the traced Polyakov loop (PL) as order parameter, to construct effective PL models, usually, the polynomial models~\cite{Pisarski:2001pe,Ratti:2006gh,Huang:2020crf,Kang:2021epo} also see review~\cite{Fukushima:2017csk}. Another line is underlined by the Haar measure, which gains great success in the $SU(3)$ case, even incorporating dynamic quarks~\cite{Mocsy:2003qw,Fukushima:2003fw,Roessner:2006xn,Fukushima:2008wg}. For $SU(3)$ only, both types of model can describe the deconfinement phase transition and as well the thermodynamics, at least in the semi-QGP region. However, when we try to extend them to general $SU(N)$ cases, we encounter some difficulties. The Haar-type model is shown to be inconsistent with the above large $N$ scaling law~\cite{Kang:2021epo} and moreover, it can not be handled for very large $N$. The polynomial model proposed in Ref~\cite{Huang:2020crf,Kang:2021epo} utilizes the competition among  terms with designed powers and signs to realize the deconfinement phase transition. Since it basically respects just a $Z_2$ symmetry and thus works for any $N$, even including $N=2$. The matrix models, inspired by the property of perturbation potential, instead  ~\cite{Meisinger:2001cq,Meisinger:2003id,Dumitru:2010mj,Dumitru:2012fw} treat the eigenvalues of the thermal Wilson line as fundamental variables, which may provide a feasible way to understand the behavior of the hot PYM system for all $N$. Largely speaking, these models are phenomenally oriented, lacking a more profound basis to derive the shape of the thermodynamic potential.

As long as only thermodynamics is concerned, the quasi-particle model (QPM) is even more  attractive. It is a statistical model where the gluons are assumed to develop a  temperature-dependent mass, due to the non-perturbative interaction with the thermal environment. This picture is strongly supported by the hard-thermal-loop perturbation theory at high-temperature regions~\cite{HTLpt}. It can successfully explain the thermodynamics of the hot $SU(N)$ PYM system from $T_c$ to the SB limit  \cite{Goloviznin:1992ws,Peshier:1994zf,Peshier:1995ty,Gorenstein:1995vm}. Later, taking into account the temporal gauge field background $A_0$ brings a difference~\cite{Meisinger:2003id} and opens the possibility to describe both thermodynamics and the deconfinement phase transition at the same time. But most studies of the interplay between quasigluon and background focus on the modification to the pressure of hot PYM, the critical non-perturbative dynamics driving the deconfinement phase transition, says the Haar measure term, is still added by hand and external to the QPM picture~\cite{Sasaki:2012bi,Ruggieri:2012ny,Islam:2021qwh}. This might be contradictory to the spirit of QPM, where most of the non perturbation interaction has already been ``absorbed" into the quasiguon mass.

Although not a following study of QPM, the massive PYM model~\cite{Curci:1976bt,Reinosa:2014ooa,vanEgmond:2021jyx} shares a similar philosophy with QPM, and it also assumes that the effective gluon mass parameter simply encodes the non-perturbative effects. Then, the effective potential for the temporal background can be derived, instead of added by hand, at one loop level or even beyond~\cite{Reinosa:2014zta,Reinosa:2015gxn}. This approach realizes the inverted Weiss potential, attributed to the enhanced ghost contribution, as the mechanism for the deconfinement phase transition. Surprisingly, for $N=2$, the resulting effective potential indeed predicts a crossover instead of FOPT. 

%with QPM, although the formal is not a following study of QPM. Both models share the philosophy that the strongly coupled system can be studied in a perturbative way, with the effective gluon mass parameter simply encoding the non perturbative effects. 

Thus, it is tempting to marry QPM with the massive PYM. The original massive PYM model~\cite{Reinosa:2014ooa} just takes a constant quasigluon mass, and now we generalize it to have temperature dependence, which is in line with the framework of hard-thermal-loop perturbation theory and may serve as a quantum field basis for the QPM. We find that the resulting one-loop effective potential indeed can successfully  describe both the deconfinement phase transition and the thermodynamics of the hot PYM system, for any color number $N$. Our study is helpful to understand the deconfinement phase transition in cosmology.

The paper is organized as follows: We give a short review of the QPM in Section II and then goes to the generalized massive PYM according to the QPM in Section III, where we derive the effective potential at one loop and investigate the deconfinement phase transition with the assumption of uniform eigenvalue distribution, which reduces the potential to one-dimension. In Section IV we study thermodynamics from the critical temperature to the SB limit, fitting the quasigluon mass by lattice data via machine learning. Conclusions and discussions and as well as the appendix are cast in the remaining two sections.

% model to describe both the de-confinement phase transition and the thermodynamics quantities. We will combined the QPM and the matrix model to build a unified frameworks for all color number $N$. To do that we will start from the first principle lagrangian with the QPM mass to get the effective potential. By using the assumption of uniform eigenvalue distribution in matrix model which can reduce the potential into one-dimensional case, for the first time, we are able to find that the effective quasi-particle mass will have a unified behavior for different $N$. We will derive a simple analytical expression of the effective potential around the critical temperature. This will be helpful to discuss confinement phase transition in cosmology in the future. We also provide a modified version of this model by considering the dressing propagator of the gluons. The modified model contain more parameter and have the potential to deal with more detailed problems.

\section{Quasigluon: From HTL to $T_c$}

For thermal gauge theories at high temperature, the classical solution should not be described by the gluonic states without mass but with mass, which stems from the plasma effects such as the screening of electric fields and Landau damping. 
The hard-thermal-loop perturbation theory (HTLpt)~\cite{HTLpt}, which is a reorganization of the perturbation series and can take into account the plasma effects consistently. It is found that at NNLO, the hot gluon plasma can be well described by weakly coupled quasigluons down to $~(2-3)T_c$~\cite{Andersen:2009tc,Andersen:2010wu}.

Within the HTLpt, the transverse quasigluon in the QCD medium  follows the dispersion equation 
\begin{equation}
w^2-k^2-\Pi_t^{*}(w,k)=0,
\end{equation}
where $\Pi_t^{*}(w,k)$ is the transverse self-energy for the hot gluons, having weak momentum dependence but strong temperature dependence. At leading order, it is given by~\cite{HTLpt}
\begin{equation}
\Pi_t^{*}(w,k)=\f{N}{6}g^2T^2,
\end{equation}
with $g$ the gauge coupling. The gluon quasiparticles mainly propagate on shell.

For even lower temperature, the magnetic/nonperturbative effects become important. But it is tempting to pursue the possibility that even down to $T_c$, the plasma can still be described by an ideal gas of ``massive" noninteracting ``gluons", where the strong interactions between gluon and the in-medium have been ``absorbed", at least partially, into the quasigluon mass. Following this line, the authors of Ref.~\cite{Goloviznin:1992ws,Peshier:1995ty} explained the lattice QCD thermodynamics near $T_c$ via a simple quasiparticle model (QPM) inspired by the above HTL quasiparticle. And to naturally match the HTLpt quasiparticle at high $T$, they simply consider such a QPM with quasigluon mass squared~\footnote{The quasigluon mass is determined by the pole of gluon  self-energy in the complex momentum plane, but the exact location is hampered by the non-perturbative effect.}
\begin{equation}\label{QPM:mass}
M_g^2(T)=\f{N}{6}G^2(T)T^2,\quad  G^2(T)=\f{48\pi^2}{11N\log\L\f{T}{T_c/\ld}+\f{T_s}{T_c}\R^2}
\end{equation}
which parameterizes the deviation from HTLpt quasiparticle via the parameter $T_s$ and $\lambda$. Other form of $M_g(T)$ is possible, for instance the one in Ref~\cite{Schneider:2001nf}.

%with $M_g(T)$ the effective mass produced by the interaction between thermal gluons and hot bath background $f_B$. So, this mass may be related to the 

%the effective mass produced by the interaction between thermal gluons and hot bath background $f_B$. So, this mass may be related to the 
%found that using the hard-thermal-loop perturbation theory (HTLpt), which is a reorganization of the perturbation series and can take into account the plasma effects such as the screening of electric fields and Landau damping. 

%In the gauge invariant HTL in Ref.~\cite{Braaten:1991gm}, 

%Quasiparticle, putatively the collective excitations of a strongly coupled plasma, is the first (and natural) model used to explain QCD thermodynamics over a wide temperature~\cite{Schneider:2001nf}. In this picture, it was assumed that the system of interacting massless gluons, at $T>T_c$, can be effectively represented as an ideal gas of "massive" noninteracting "gluons"~\cite{Goloviznin:1992ws,Peshier:1995ty}. 

Then, such a pool of ideal quasi gluon gas, assumed to respect the Bose distribution $f_B$, has pressure
\begin{equation}\label{pressure:old}
p(T)=\f{g(T)}{6\pi^2}\int_0^\infty f_B(E_k)\f{k^4}{E_k} dk-B(T),
\end{equation}
with $E_k=\sqrt{k^2+M_g^2 (T)}$. Owing to the temperature dependence of mass, the self-consistent thermodynamic relation for ideal gas, namely the Gibbs-Duhem relation, $\epsilon+p=sT$ with $s=\partial p/\partial T$ is violated. Including $B(T)$ can solve this problem~\cite{Gorenstein:1995vm}. It is not independent and is determined by $M_g(T)$, up to a bag constant. Surprisingly, this simple QPM is capable of reproducing the quenched QCD or $SU(3)$ PYM lattice data in the whole region above $T_c$~\cite{Peshier:1995ty}. Study for Other $N=4,5,6$ is presented in Ref.~\cite{Castorina:2011ra,Castorina:2011ja}. It is common that, in order to reduce the contribution of quasi gluons near the critical temperature, a very large quasi gluon mass is usually required. 

%For temperature $T\gtrsim 3T_c$. HTL resummation perturbation theory does support that QGP is a weakly interacting quasiparticle gas. 

However, the original QPM is just a statistical model and thus can not explain the order of $SU(N)$ phase transition. The latter is supposed to be understood in the framework of Landau phase transition: Find a proper order parameter $\eta$ and construct a (coarse-grained) Landau free energy as a function of the order parameter, and then one can study the order of phase transition by surveying its ground state. In studying the deconfinement phase transition of $SU(N)$ PYM, the Polyakov loop (PL) associated with the center symmetry $Z_N$ is identified with $\eta$; it is defined as 
 $l_N={\rm tr} \hat{L}_F/N$, the traced thermal Wilson line in the fundamental representation 
\begin{equation}\label{}
\hat L_F={\cal P}e^{ig\int_0^\beta A^a_4(x,t)t^adt},
\end{equation}
with ${\cal P}$ denoting path ordering and $t^a$ the generators of the fundamental representation for $SU(N)$.  

%PL models have been proposed to study deconfinement transition~\cite{}, and they are additionally required to fit the thermodynamics over a wide temperature, but usually mission impossible~\cite{}, which may be expected since the Landau free energy is supposed to be valid only around the temperature of phase transition $T_c$. 

%, a $SU(N)$ group element,

\section{Quasiparticles move in the PL background}

So, it is a natural idea to combine quasi-particle model with PLM, to study the deconfinement phase transition dynamics and thermodynamics simultaneously~\cite{Meisinger:2003id,Ruggieri:2012ny}~\footnote{This idea originated from an earlier work~\cite{Meisinger:2001cq}, although there the authors have not introduced quasigluon explicitly yet. }. In such models, quasigluons moving in the PL background generate thermodynamic potential which depends on the PL in the adjoint representation $\hat{L}_A$~\cite{Meisinger:2003id}:
\begin{equation}\label{potential}
\Omega_{\rm QG}(\hat L_A,T)=2T{\rm tr}\int\frac{d^3\vec{p}}{(2\pi)^2}\log(1-\hat{L}_A e^{-E_g/T}).
\end{equation}
It is a phenomenological generalization to the usual Weiss potential~\cite{Weiss:1980rj} for the fundamental gluons to quasigluons, by replacing $|\vec{p}|$ with $E_g=\sqrt{M_g(T)^2+p^2}$. Later, we will derive a similarity grounded on the QFT, but with a remarkable difference. The quasigluons dominate thermodynamics in the high-temperature region, where $M_g\ll T$ and $\hat L_A\ra 1$, explain the blackbody behavior. At the lower temperature, typically below $2T_c$, the decreasing PL combined with the increasing $M_g(T)$, is capable of explaining the deviation from the blackbody spectrum  towards $T_c$~\cite{Meisinger:2003id, Sasaki:2012bi,Ruggieri:2012ny,Alba:2014lda,Islam:2021qwh}. 

But that's all. We can't expect this part to give the deconfinement phase transition at the same time, which needs additional interaction, such as the van der Monde determinant interaction~\cite{Meisinger:2003id, Sasaki:2012bi,Lo:2021qkw,Islam:2021qwh}. In this article, we follow another line proposed in Ref.~\cite{Reinosa:2014ooa}, which enables us to study the non-perturbative PT in the perturbative approach; in their philosophy, non-perturbation effects are encoded in the gluon mass, in line with the QPM picture. In the following, we will first present an effective model, which is a slight generalization to that in Ref.~\cite{Reinosa:2014ooa}. Then, we reproduce the effective potential Eq.~(\ref{potential}) as well as the confining potential from the model through the leading order thermal correction.

%near the critical point and account for PT, while And they interplay around $1.5T_c$. Then, quasiparticles moving in the nontirivial PL background could successfully explain the thermodynamics above $T_c$~\cite{Meisinger:2003id}. 

%In principle, we should calculate this potential in the thermal QFT framework. However, it is nontrivial to move from the original QPM, a statistical model, to the thermal QFT. 

\subsection{Effective model for quasigluon above $T_c$}

The model in Ref.~\cite{Reinosa:2014ooa} quantizes PYM in the background field gauge formalism, including massive fluctuations. Then, the Faddeev-Popov gauge-fixed Lagrangian reads
\begin{equation}\label{def:la}
\mathcal{L}=-\frac{1}{2g^2}{\rm tr}(F_{\mu\nu}F^{\mu\nu})+\bar{D}_\mu \bar{c}^a D^\mu c^a+ih^a \bar{D}_\mu\hat{A}^{\mu,a}+\f{1}{2}M^2_g(T)\hat A^a_{\mu} \hat A^{a,\mu},
\end{equation}
where $c,\bar c$ and $h$ are the Ghost fields, real Nakanishi-Lautrup field, respectively. We have split the gauge field $A_\mu$ as $A_\mu=\bar{A}_\mu+\hat{A}_\mu$ with $\hat{A}_\mu$ the massive fluctuations. The background $\bar{A}_\mu$ is restricted to merely have the constant temporal component, $\bar{A}_\mu=\bar{A}_0\delta_{0\mu}$, for the sake of preserving invariance of the PYM system, under both temporal and spatial translations and spatial rotations at finite $T$. The covariant derivative acting on $\phi=(c,\bar c, h,\hat A_\mu)$ is defined as 
\begin{equation}\label{}
\bar{D}_\mu^{ab}=\partial_\mu\delta^{ab}+gf^{acb}\bar A^c_\mu,
\end{equation}
where the gauge field is the background field. The above Lagrangian has implemented the Landau-DeWitt gauge $\bar D_\mu \hat A^\mu=0$. This gauge fixed PYM, including the gluon mass term, still respects the background local $SU(N)$ symmetry, with covariant derivative defined above and treating $\phi$ as adjoint matter fields.

In the effective model specified by Eq.~(\ref{def:la}), the gluon mass is not originally interpreted as quasigluon mass. Instead, it is regarded as a gauge fixing parameter, to further remove the degeneracy among the Gribov copies, whose existence may make the Faddeev-Popov procedure in the deep infrared region invalid~\cite{Gribov:1977wm}. This region is associated with the nonperturbative 
dynamics of PYM. Hence, people hope that  $M_g$ at the same time can ``absorb" strong interactions, so that some non-perturbation phenomena can be studied by the perturbation method. Such a philosophy is consistent with the QPM, and therefore it is tempting to simply identify $M_g$ as the quasigluon mass, which is reasonable at least at zero temperature. If such a formalism is consistent with the Hamilton approach which establishes a QFT basis for quasiparticle~\cite{Heffner:2012sx}, is open.

However, to explain thermodynamics, we need a temperature-dependent quasigluon mass, $M_g(T)$. This may be odd with the usual understanding of thermal mass origin in perturbative thermal QFT: The underlying Lagrangian is the same as that of $T=0$ and does not include the temperature-dependent quantity, and this kind of dependence originates from thermal correction. However, it is not strange that the Lagrangian includes a temperature dependent quantity. In fact, the HTL resummation scheme based on quasi-particle picture is just based on the effective Lagrangian including thermal mass, which gives rise to the modified propagator for the calculation of thermal corrections. Since we are extending the quasi particle picture down to near $T_c$, we should naturally include the temperature-dependent quasigluon mass term.

Therefore, as a slight generalization to the model in Ref.~\cite{Reinosa:2014ooa}, the effective Lagrangian Eq.~(\ref{def:la}) is supposed to furnish a phenomenological framework to perturbatively study deconfinement phase transition along with full thermodynamics above $T_c$.

\subsection{The Thermodynamic Potential for Quasi-Particle Model: pure gluonic part}

In this subsection, we will calculate the thermodynamic potential for the fundamental PL in a general PYM with gauge group $SU(N)$, following the textbook approach. That is to integrate out all fluctuations $\hat{A}_\mu=A_\mu-\bar{A}_\mu$ over the temporal background $\bar{A}_\mu=\bar{A}_0\delta_{0\mu}$, in the $3+1$ Eucleadian QFT. For a homogeneous background, one can always make $\bar A_0$ diagonal via some global $SU(N)$ rotation. Therefore, we can expand $\bar A_0$ in the $su(N)$ Cartan space, which is spanned by the diagonal subgroup $\{H^i\}$ $(i=1,2,...N-1)$ with $[H^i,H^j]=0$, and then $\bar A_0=\bar A_0^i H^i$ with $\bar A_0^i$ is the Cartan  coordinates. 

Let us first deal with the pure gluonic part of Eq.~(\ref{def:la}), from which one can get the quadratic Lagrangian of the fluctuation field $\hat{A}_\mu$
\begin{equation}\label{QE}
\mathcal{L}^{(2)}=-\frac{1}{2}\hat{A}_{\mu}^a[\delta_{ab}g^{\mu\nu}\partial^2-f_{abc}(\partial^\nu\bar{A}^{\mu,c}+2g^{\mu\nu}\bar{A}^c_\rho\partial^\rho)+f_{acd}f_{cbe}g^{\mu\nu}\bar{A}^d_\rho\bar{A}^{\rho,e}+2f_{abc}\bar{F}^{\mu\nu,c}]\hat{A}^b_\nu.
\end{equation}
It can be written as the following
\begin{equation}
\mathcal{L}^{(2)}=\frac{1}{2}\hat{A}_\mu^a(D^{-1})_{ab}\hat{A}^{\mu,b},
\end{equation}
with the operator defined as 
\begin{equation}\label{D-1}
(D^{-1})_{ab}=\delta_{ab}(p^2+M_g^2)+2i\sum_{i}f_{abi}\bar{A}^i_0 p_0-\sum_{i,j}f_{aci}f_{cbj}\bar{A}^i_0\bar{A}^j_0.
\end{equation}
The last term denotes the mass of the fluctuations (explained as the quasigluons) from the  temporal background, and hence the background field will obtain a thermodynamic potential from the plasma of quasigluons.

Before we calculate this potential, let's deal with the propagator, diagonalizing the fluctuations in the  color space through a unitary transformation.  Then, the diagonal propagators takes the form of
\begin{equation}\label{inverse:p}
\tilde{D}^{-1}_{aa}(p)=(p_0-A_a)^2+|\vec{p}|^2+M_g^2.
\end{equation}
Following the standard approach of path integral, one can get the generating function ($Z$ below should be understood as $Z^I$, the gluonic part contribution, but for the sake of simplicity, we ignore the superscript, which we believe will not cause ambiguity)
\begin{equation}\label{effective action}
\log Z=\frac{1}{2}\log\det\left[-\frac{\delta^2\mathcal{L}^{(2)}}{\delta A\delta A}\right]=\frac{1}{2}\log\det\left[D^{-1}\right]=\frac{1}{2}\log\det\left[\tilde{D}^{-1}\right].
\end{equation}
where we have used the property that unitary transformation does not change determinate. Then, using the trick that $\log \det A={\rm Tr} \log A$ we get
\begin{equation}
\log \det [\tilde{D}^{-1}]={\rm Tr} \log [\tilde{D}^{-1}].
\end{equation}
``{\rm Tr}" is the trace over the functional propagator operator, and can be split into two parts: a function trace over momentum space and a color space trace denoted by ``${\rm tr_c}$", explicitly, 
\begin{equation}
{\rm Tr} \log [\tilde{D}^{-1}]={\rm tr_c}\int\frac{d^4p}{(2\pi)^4} \log [\tilde{D}^{-1}(p)].
\end{equation}
In order to get the finite temperature potential, one can discretize the energy by $p_0\rightarrow \omega_n=2i\pi nT$ and transform $\bar{A}_0\rightarrow -i\bar{A}_4$, obtaning 
\begin{equation}\label{Eac}
\log Z=2 V{\rm tr_c}\int\frac{d^3\vec{p}}{(2\pi)^2}\sum_{n=-\infty}^\infty\log[\tilde{D}^{-1}_{aa}(\omega_n,|\vec{p}|)],
\end{equation}
where $V$ is the space volume and $2=\f{1}{2}\times 4$ with 4 denoting the multiplicity from the four components of $A_\mu$. From Eq.~(\ref{inverse:p}), the structure of the propagator $\tilde{D}^{-1}(\omega_n,|\vec{p}|)$ takes the form of
\begin{equation}
\tilde{D}^{-1}_{aa}(\omega_n,|\vec{p}|)=(\omega_n-A_a)^2+|\vec{p}|^2+M_g^2,
\end{equation}
where $A_a$ is a linear combination of the background $\bar A^i_4$, with coefficients determined by the structure constant, but we do not find a general expression for any $N$ yet. As a matter of fact, the concrete expression is not important in our discussion, since later we will switch to a parameterization of the background which is independent of $A_a$. Anyway, in Appendix.~\ref{SU(4)}, we present the details of our calculation for $SU(4)$, and the procedure applies to other values of $N$.

The summation of the thermal excitation modes $n$ can be done explicitly using a trick in Appendix.~\ref{GFcomputation1}. And finally, the generating function can be compactly written as
\begin{equation}\label{weiss1}
\log Z=4V{\rm tr_c}\int\frac{d^3\vec{p}}{(2\pi)^2}\log(1-\hat{L}_A e^{-E_g/T}),
\end{equation}
where $\hat{L}_A$ is expressed in terms of background field $\bar{A}_\mu$, and it is nothing but the PL in the adjoint representation. For instance, in $SU(3)$ it is given by
\begin{equation}\label{LA}
\begin{split}
\hat{L}_A=\diag[&1,1,e^{i\bar{A}^3_4/T},e^{-i\bar{A}^3_4/T},e^{i(\bar{A}^3_4+\sqrt{3}\bar{A}^8_4)/2T},
\\
&e^{-i(\bar{A}^3_4+\sqrt{3}\bar{A}^8_4)/2T},e^{i(\bar{A}^3_4-\sqrt{3}\bar{A}^8_4)/2T}
,e^{-i(\bar{A}^3_4-\sqrt{3}\bar{A}^8_4)/2T}],
\end{split}
\end{equation}
where we have written it in terms of the original background. It is seen that the temporal background behaves as an imaginary chemical potential. Eq.~(\ref{weiss1}) yields the effective potential ${\cal V}_{eff}=\f{T}{V}\log Z$ which almost recovers the generalized Weiss potential given in Eq.~(\ref{potential}), up to the coefficient. But the ghost contribution, which will be included in the following subsection, will result in a substantial deviation related to the deconfinement potential.

Eq.~(\ref{LA}) demonstrates the general structure of thermal Wilson line in the adjoint representation, i.e., its elements are organized such that it can be rewritten  in terms of the eigenphases of the fundamental thermal Wilson line~\cite{Lo:2021qkw} \begin{equation}\label{}
\hat{L}_F={\cal P}e^{ig\int_0^\beta A_4(x,t)dt}\ra \diag[e^{i2\pi q_1},e^{i2\pi q_2},...,e^{i2\pi q_N}],
\end{equation}
by virtue of the parameterization of background 
$\bar A_4=2\pi/(g\beta){\rm diag}(q_1,q_2,...,q_N)$, with the real $q_i$ satisfying the constraint $\sum_{i=1}^Nq_i=0$. As a phase factor, it is sufficient to work in the interval $0\leq q_i\leq 1$. And now, 
\begin{equation}\label{}
 \hat{L}_A =\diag[1,1,...,1,e^{i2\pi q_{ij}},...,e^{-i2\pi q_{ij}}],
\end{equation}
where the $N-1$ ``1" corresponds to the Cartan part, while the $N(N-1)/2$ pairs of $q_{ij}\equiv q_i-q_j$ with $N\geq i>j\geq 1$ corresponds to the non-Cartan part. The above form is more convenient and will be adopted hereafter. Then $1-\hat{L}_A e^{-E_g/T}=\diag(1- e^{-E_g/T},...,1- e^{i2\pi q_{ij}-E_g/T},...,1- e^{-i2\pi q_{ij}-E_g/T})$. We also define 
\begin{align}\label{omega}
\Omega(q_{ij},M_g)&\equiv 2T\int\frac{d^3\vec{p}}{(2\pi)^2}\log(\det (1-\hat{L}_A e^{-E_g/T}))\\
\nonumber
&=2T\sum_{i,j=1}^N\L 1-\frac{\delta_{ij}}{N}\R\int \frac{d^3\vec{p}}{(2\pi)^3} \log(1-e^{- E_g/T}e^{2\pi i q_{ij}}),
\end{align}
where in the second line we use $q_{ij}=-q_{ji}$ and allow $i=j$, to write the summation compactly~\footnote{Actually, if we instead adopt the ladder basis for $su(N)$ and start from the eigenphase parameterization of $\bar A_4$~\cite{Fukushima:2017csk}, the above expression can be explicitly obtained from the covariant derivative whose background dependent term reads $[\bar A_4, \hat A_\mu]\sim (q_i-q_j)\delta_{\mu 4}$.}. In this notation, the gluonic part contribution to the effective potential ${\cal V}_{eff}$ is $2\Omega(q_{ij},M_g)$.

In the following, we present two important expansions of this potential, the low temperature and the high temperature expansion. Both will be used in the later discussions.

\subsubsection{Low temperature expansion}
 
In the QPM, it is found that the fitted $M_g(T)/T$ is sufficiently large at least around $T_c$, hence one has $E_g/T>M_g/T\gtrsim {\cal O}(1)$. We will find this is also true in our model from a full numerical study, which enables us to make a low temperature expansion for the effective potential around $T_c$. This leads to an analytical expression, which is useful in the phase transition analysis. First, we expand the logarithm in $\Omega(q_{ij},M_g)$, retaining $\hat L_A$, 
\begin{equation}\label{}
\Omega(\hat L_A,M_g)=-\f{T}{\pi^2}\sum_{n=1}^\infty\f{1}{n} {\rm tr}(\hat{L}_A)^n \int  p^2 e^{-nE_g/T} dp.
\end{equation}
Then we substitute $p=M_g\sinh t$ to get 
\begin{align}\label{}
\Omega(\hat L_A,M_g)=&-\f{TM_g^3}{\pi^2}\sum_{n=1}^\infty\f{1}{n} {\rm tr}(\hat{L}_A)^n \int_0^\infty \cosh t\sinh^2 t   e^{-n(M_g/T)\cosh t} dt
\end{align}
Now we use the following trick to rewrite the integral as
\begin{equation}\label{}
\Omega(\hat L_A,M_g)= \f{T^2M_g^3}{\pi^2}\sum_{n=1}^\infty\f{1}{n^2} {\rm tr}(\hat{L}_A)^n \f{d}{dM_g}\int_0^\infty  \sinh^2 t   e^{-n(M_g/T)\cosh t} dt,
\end{equation}
where the integral can be done explicitly, $\sim K_1(x)$, and we finally arrive
\begin{align}\label{LTE}
\Omega(\hat L_A,M_g)
% =& \f{T^2M_g^3}{\pi^2}\sum_{n=1}^\infty\f{1}{n^2} {\rm tr}(\hat{L}_A)^n \f{d}{dM_g}\left[\f{T}{nM_g}K_1(nM_g/T)\right].\\
% \nonumber
=- \f{T^2M_g^2}{\pi^2}\sum_{n=1}^\infty\f{1}{n^2} {\rm tr}(\hat{L}_A)^n K_2(nM_g/T).
\end{align}
with $K_i(x)$ the modified Bessel function of the second kind, of order $i$. For $M_g/T$ moderately larger than 1, the leading order is a good approximation.

\subsubsection{High temperature expansion}

At the high temperature limit, where $1\gg M_g/T$, one can find a simple analytic expression of $\Omega(q_{ij},M_g)$; see also Ref.~\cite{Meisinger:2001fi} for the complete high temperature expansions beyond the leading term. To that end, we again expand the logarithmic function:
\begin{equation}
    \Omega(q_{ij},M_g)=-\frac{T^4}{\pi^2}\sum_{i,j=1}^N\L 1-\frac{\delta_{ij}}{N}\R\int_0^\infty dx x^2 \sum_{n=1}^\infty \frac{1}{n} e^{2\pi n i q_{ij} } e^{-\sqrt{n^2(x^2+\beta^2 M_g^2)} }.
\end{equation}
Expand this expression according to $\beta M_g$, and we get 
\begin{equation}
\begin{split}
    \Omega(q_{ij},M_g)
    &=-\frac{T^4}{\pi^2}\sum_{i,j=1}^N\L 1-\frac{\delta_{ij}}{N}\R\int_0^\infty dx x^2 \sum_{n=1}^\infty \frac{1}{n} e^{2\pi n i q_{ij} } \left[ e^{-nx}-\beta^2 M_g^2\frac{n e^{-nx}}{2x} \right]+\mathcal{O}(\beta^2M_g^2).
  \end{split}
\end{equation}
The summation over $n$ in the first term is straightforward, while in the second term, with one more $``n"$ factor from the Taylor expansion, can be done as the following,
\begin{equation}
\begin{split}
   \sum_{n=1}^\infty \frac{1}{n} n e^{-nx} = -\L\sum_{n=1}^\infty \frac{1}{n}e^{-nx } \R'=-\f{d }{dx}\log\L 1-e^{-x}\R.
\end{split}
\end{equation}
Similar operations can be generalized to higher order of $n$, leading to higher derivative to $\log\L 1-e^{-x}\R$. Now, the first two terms are summed to
\begin{equation}
\begin{aligned}
    \Omega(q_{ij},M_g)
    &=-\frac{T^4}{\pi^2}\sum_{i,j=1}^N\L 1-\frac{\delta_{ij}}{N}\R\int_0^\infty dx x^2  [\log\L 1-e^{-x+2\pi i q_{ij}}\R&
    \\
    &+ \frac{\beta^2M_g^2}{2x}\f{d}{dx}\log\L 1-e^{-x+2\pi i q_{ij}}\R].
\end{aligned}
\end{equation}
The result of these integration are just two Polylogarithm function
\begin{equation}
\begin{split}
    \Omega(q_{ij},M_g)=-\frac{T^4}{\pi^2}\sum_{i,j=1}^N\L 1-\frac{\delta_{ij}}{N}\R\left[2 Li_4(e^{2\pi i q_{ij}}) -\frac{\beta^2M_g^2}{2} Li_2(e^{2\pi i q_{ij}}) \right]+\mathcal{O}(\beta^2 M_g^2).
\end{split}
\end{equation}
Since $q_{ij}=q_i-q_j=-q_{ji}$ we can rewrite this expression into an analytic form by Jonquière's inversion formula
\begin{equation}
\begin{split}
    Li_n(e^{2\pi i x})+(-1)^n Li_n(e^{-2\pi i x})=-\frac{(2\pi i)^n}{n!}B_n(x),
\end{split}
\end{equation}
where $B_n$ is the Bernoulli polynomials. In our case, we can find that
\begin{equation}
\begin{split}
    &Li_2(e^{2\pi i q_{ij}})+ Li_2(e^{2\pi i q_{ji}})=2\pi^2 B_2(q_{ij}),
    \\
    &Li_4(e^{2\pi i q_{ij}})+ Li_4(e^{2\pi i q_{ji}})=-\frac{2\pi^4}{3} B_4(q_{ij}).
\end{split}
\end{equation}
Substitute this formula into our expression to get
\begin{equation}
\begin{split}
    \Omega(q_{ij},M_g)=\frac{1}{\pi^2}\sum_{i\geq j=1}^N\L 1-\frac{N-1}{N}\delta_{ij}\R\left[\frac{4\pi^4}{3\beta^4}B_4(q_{ij})+\frac{M_g^2\pi^2}{\beta^2}B_2(q_{ij}) \right]+\mathcal{O}(\beta^2 M_g^2),
\end{split}
\end{equation}
We can write it into a more simple form
\begin{equation}\label{HTE}
\begin{split}
    \Omega(q_{ij},M_g)=\frac{1}{\pi^2}\sum_{i,j=1}^N\L 1-\frac{\delta_{ij}}{N}\R\left[\frac{2\pi^4}{3\beta^4}B_4(|q_{ij}|)+\frac{M_g^2\pi^2}{2\beta^2}B_2(|q_{ij}|) \right]+\mathcal{O}(\beta^2 M_g^2).
\end{split}
\end{equation}
In the massless limit only $B_4$ is present.

\subsection{The Thermodynamic Potential: gauge-fixed part \& Phase transition}

In this subsection, we include the contribution from the gauge-fixed part to the thermodynamic potential, and then study how first order deconfinement phase transition occurs due to the ghost contribution~\cite{Reinosa:2014ooa}. 

\subsubsection{Infrared ghost domination}

The ghost contribution is similar to that from the gluons because it is also in the adjoint representation. However, there are two key differences, which enable the contribution of the thermodynamic potential from the ghost fields to successfully trigger the deconfinement phase transition. First, the ghost fields belong to Grassmann fields, and thus there is a minus sign relative to the gluon contribution. Second, the ghosts are still massless since the Lattice data does not show that the correlators of ghost develop a massive pole. Moreover, we have to take into account the contribution from the gauge-fixing term. To deal with this term, we should do the quadratic partition between $\hat A^a$ and field $h^a$, to get two quadratic terms with the mixing term eliminated; the details can be found in Appendix~\ref{GFcomputation} or in the textbook~\cite{Reinosa:2020mnx}. The final result of the total effective potential is given by
\begin{equation}\label{def:EP}
\begin{split}
   {\cal V}_{eff}(M_g)=\f{3}{2}\Omega(M_g)-\f{1}{2}\Omega(M_g=0),
\end{split}
\end{equation}
The above result is in the Landau-DeWitt gauge, and as usual, the effective potential is gauge-dependent.

Without a quasi-gluon mass, the ghost contribution cancels the nonphysical gluonic contribution and then the potential fails to admit a phase transition. On the contrary, the presence of $M_g$ makes the enhanced ghost contribution (relative to gluon contribution) dominate the potential at low temperature, realizing the inverted Weiss potential as the confining mechanism.

%RQuasigluons moving in the PL background potentially account for the thermodynamics, however, they fail in accounting for the confinment PT. It is noticed that the effective potential Eq.~(\ref{potential}) may be regarded as the deconfining part of the total potential to explain the correct behavior of deconfinement phase transition, since it prefers the spontaneously broken of $Z_N$. A simple observation is that, at the leading order of perturbation theory, the ghost contribution to the effective potential is very similar to that of the gluon's contribution in the limit of zero gluon mass, but taking a opposite sign~\cite{Reinosa:2014ooa}. Thus, the ghosts are supposed to provide a confining potential.  

%But in the spirit of Kugo-Ojima~\cite{} and Gribov-Zwanziger (KOGZ) scenario of confinment, the gluon mass term may encode information of infrared dynamics, and thus provides a framework, namely the massive Faddeev-Popov Lagrangian, to describe confinement phenomenologically. 

%it is of interest to pursue that goal by possible to 

% Following the same method in the previous section, one can compute the effective potential for Faddev-Popov Lagrangian in Landau-DeWitt gauge and find the effective thermal potential is given by~\cite{Lo:2021qkw}. 
% \begin{equation}\label{def:EP}
% \begin{split}
%   V_{eff}=\f{3}{2}\Omega(M_g)-\f{1}{2}\Omega(M_g=0),
% \end{split}
% \end{equation}

Because we are dealing with the phase transition where $T\rightarrow T_c$ and $M_g/T\gg1$, it is fair for us to expand the first term by low temperature expansion Eq.(\ref{LTE}). The second term is just the zero mass limit of Eq.(\ref{HTE}) and thus high temperature expansion applies. Combining the above information, we can get the following analytic form of this effective potential
\begin{equation}\label{def:EP}
\begin{split}
    \mathcal{V}_{eff}
    &=3T\int \frac{d^3\vec{p}}{(2\pi)^3}{\rm tr}\log(1-\hat{L}_A e^{-\frac{E_g}{T}})-T\int \frac{d^3\vec{p}}{(2\pi)^3}{\rm tr}\log(1-\hat{L}_A e^{-\frac{|\vec{p}|}{T}})
    \\
    &\simeq-\frac{3T^4}{2\pi^2}(\frac{M_g}{T})^2 K_2(M_g/T) {\rm tr}(\hat{L}_A)-\frac{1}{2\pi^2}\sum_{i,j=1}^N(1-\frac{\delta_{ij}}{N})\Big[\frac{2\pi^4}{3\beta^4}B_4(q_{ij})\Big].
\end{split}
\end{equation}
The first term can be translated to a  function of PL, by using the identity ${\rm tr}\hat{L}_A={\rm tr}\hat{L}_F{\rm tr}\hat{L}_F^\dagger-1$. Nevertheless, the ghost term contains $N-1$ independent variables $q_i$, rather than merely the trace part of $\hat L_F$. Hence, usually, one has to deal with a multi-dimensional field space, case by case.

A way to reduce the potential to the  one-dimension problem is assuming the  uniform eigenvalue distribution, i.e., $q_{ij}=\frac{i-j}{N}r$. It is automatically true for $N=2,3,4$ with the number of independent eigenphases less than 4, but it is merely an ansatz for the even higher $N$. Such an ansatz has been adopted in Ref.~\cite{Dumitru:2012fw}, and is shown to work well. The ansatz is based on the observation that the confining vacuum, which is center symmetric, is characterized by the uniform eigenvalue distribution; dynamically, the distribution is a result of the eigenvalue repulse from the confining potential which involves the difference between eigenvalues $q_i$~\cite{Ivanov:2004gq,Pisarski:2006hz,Dumitru:2012fw}. Furthermore, it is conjectured that the transition from the deconfining vacuum to the confining vacuum takes the shortest path, a straight line connecting the origin and the confining vacuum~\cite{Dumitru:2012fw}.~\footnote{This is not the very precise statement. At high $T$, the origin namely $A_0=0$ is the deconfining vacuum. As $T$ decreases to near $T_c$, it moves away from the origin to the configuration still characterized by the uniform eigenvalue distribution. In the effective matrix model, this is shown to be a good approximation even for the relatively small $N$~\cite{Dumitru:2012fw}.} Then, we get the analytic potential for any color number
\begin{equation}
\begin{split}
    \mathcal{V}_{eff}
    &\simeq-\frac{3T^4N^2}{2\pi^2}(\frac{M_g}{T})^2 K_2(M_g/T)l_N^2-\frac{2\pi^2T^4}{3}\sum_{i=1}^N(N-i)B_4(\frac{i}{N}r)+f(N,T),
\end{split}
\end{equation}
where $f(N,T)$ is a function does not depend on the order parameter, only relevant to thermadynamics. We can easily carry out this summation and find that
\begin{equation}\label{EffP}
\begin{split}
    \mathcal{V}_{eff}
    &\simeq -\frac{N^2}{2}\Bigg(\frac{3T^4}{\pi^2}(\frac{M_g}{T})^2 K_2(M_g/T)l_N(r)^2
    \\
    &\ \ \ \ \ \ \ \ +\frac{\pi^2T^4}{45}\Big[(-1+r)^2(-1-2r+2r^2)-\frac{5(-1+r)^2r^2}{N^2}+\frac{r^3(-4+3r)}{N^4} \Big]\Bigg)+f(N,T),
\end{split}
\end{equation}
which is somehow a hybrid of the PL model and matrix model.

% o rewrite the  in terms of PL then we find
% \begin{equation}
% \begin{split}
%     \mathcal{V}_{eff}
%     &\simeq-\frac{3T^4N^2}{2\pi^2}(\frac{M_g}{T})^2 K_2(M_g/T)(l_N^2-\frac{1}{N^2})-\frac{1}{2\pi^2}\sum_{i,j=1}^N(1-\frac{\delta_{ij}}{N})\Big[\frac{2\pi^4}{3\beta^4}B_4(|q_{ij}|)\Big].
% \end{split}
% \end{equation}

Usually, $l_N={\rm tr} \hat{L}_F/N$ as a function of $s$ is complicated. By definition, we can find that
\begin{equation}
\label{ls}
l_N(r)=\left\{
\begin{aligned}
\frac{1}{N}\bigg(1+2\sum_{i=1}^{\frac{N-1}{2}}\cos(2\pi\frac{i}{N}r)\bigg) & , &  N\ {\rm is\ odd}, \\
\frac{2}{N}\bigg(\sum_{i=1}^{\frac{N}{2}}\cos(2\pi\frac{2i-1}{2N}r)\bigg) & , & N {\rm \ is\ even}.
\end{aligned}
\right.
\end{equation}
The summation can be implemented by writing $\cos (nx)={\rm Re}\exp{(inx)}$, translating it to the geometric series, and then we obtain the simple expression
\begin{equation}
l_N(r)=\f{1}{N}\f{\sin(\pi r)}{\sin (\pi r/N)},
\end{equation}
which holds both for odd and even $N$. In particular, when the color number approaching infinity, PL takes the limit $ {\sin(\pi r)}/({\pi r})$. 
%We plot $l_N(s)$ for different $N$ in Fig.\ref{Fig:ls}, and one can see that, when $N$ is large enough, the behavior of $l_N(s)$ hardly changes with $N$. 
% \begin{figure}[htbp]
% \centering 
% \includegraphics[width=0.6\textwidth]{l-s.pdf} 
% \caption{The trace of Polyakov loop $l$ as a function of the average eigenvalue $s$ for different color number $N$. The different color represent the different color number $N$. This function $l(s)$ has universal behavior for large $N$. } \label{Fig:ls}
% \end{figure}

%This universal behavior of $l_N(s)$ for $N>3$ allows us to generate our discussion of phase transition into any color number $N$ with very good accuracy. 

\subsubsection{Deconfinement phase transition}

Now we arrive the effective potential which can be used to study deconfinement phase transition for any color number $N$. For a sufficiently large $N$, one can simply use the following rescaled potential
\begin{equation}
\begin{split}
    \f{\mathcal{V}_{N}(r,T)}{N^2/2}
    &\simeq -\frac{3T^4}{\pi^2}\L\frac{M_g}{T}\R^2 K_2(M_g/T)\bigg[\frac{\sin(\pi r)}{\pi r}\bigg]^2+
    \frac{\pi^2T^4}{45} (r-1)^2(1+2r-2r^2),
\end{split}
\end{equation}
which is $N$-independent. To find the vacuum position of this potential, one should calculate the derivative of this potential with respect to  $r$, to solve the following tadpole equation
\begin{equation}\label{tadpole}
\begin{split}
   & \frac{6}{\pi^2}\L\frac{M_g}{T}\R^2 K_2(M_g/T)\frac{ \pi  \sin (\pi  r)  \left(
   \cos (\pi  r)-\f{1}{N}\sin (\pi  r) \cot \left(\frac{\pi 
   r}{N}\right)\right)}{N^2\sin^2(\pi r/N)}
    \\
    & \quad\quad\quad\quad \quad \quad \quad\quad \quad+\frac{\pi^2}{45}r(r-1)\Big[2(4r-5)-\frac{10}{N^2}(2r-1)+\frac{12}{N^4}r \Big]=0.
\end{split}
\end{equation}
It has two obvious solutions: 1) $r=0$, the deconfined vacuum position at high temperature; 2) $r=1$, the confining vacuum position, which is consistent with the $Z_N$ symmetry argument: The confining vacuum should preserve the $Z_N$ symmetry and then the PL value must be $l_N(r=1)=0$. In our model, this is trivially satisfied for the potential from gluons, which contributes the quadratic term $l_N^2$. However, the inverted Weiss term is not a polynomial of $l_N$, and therefore $r=1$ (namely $l_N=0$) being its extremum is nontrivial. It is attributed to the eigenvalue repulse of the potential. Eq.~(\ref{tadpole}) may also admit solutions for $r\neq 1$, the candidates for the deconfined vacuum at the lower temperature.

%Now, let us to consider the phase transition itself. Using the expression of effective potential Eq.(\ref{EffP}), one can find 

\begin{figure}[htbp] 
\centering 
\includegraphics[width=0.49\textwidth]{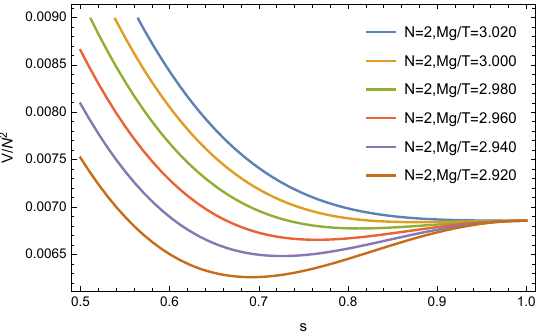} 
\includegraphics[width=0.49\textwidth]{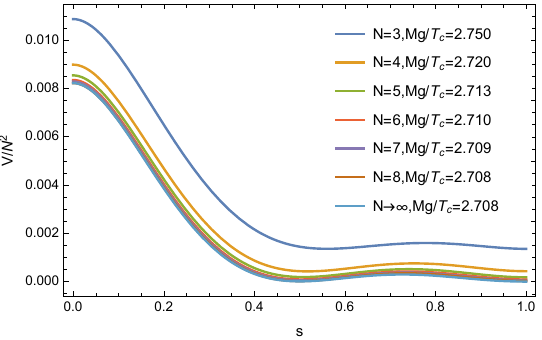} 
\caption{Left panel: The potential behavior for $SU(2)$ theory around the critical temperature. Right panel: The potential shape for $SU(N\geq 3)$ at the critical temperature.} \label{Fig:V-s}
\end{figure}
For a given $N$, the shape of the potential Eq.(\ref{EffP}) is solely determined by the single dimensionless parameter $M_g(T)/T$. 
% {\red and position of deconfinement phase $s_d$, the order of the confinement phase transition can be determined by equation group
% \begin{equation}
% \begin{split}
%     &\frac{d}{ds}\frac{V_{eff}(s,M_g(T_c)/T_c)}{T_c^4}\vert_{s=s_d}=0
%     \\
%     &\frac{V_{eff}(s_d,M_g(T_c)/T_c)}{T_c^4}=\frac{V_{eff}(1,M_g(T_c)/T_c)}{T_c^4}
% \end{split}
% \end{equation}
% where two equation can determine two parameter $s_d$ and $M_g(T_c)/T_c$ for concrete $N$. The first equation representing the vacuum position for deconfinement phase and the second the degenerating vacuum phase condition for FOPT. If one can find a non-trivial solution($s\neq 0,1$) for this equation group then the de-confinement phase transition is a FOPT but if solution do not exist then the phase transition is crossover type since there is one 1 vacuum for any $M_g(T_c)/T_c$. However those equation group for any color number $N$ is very complex and very difficult to solve even for the numerical case. Based on this, one can use a clever method to find the solution of that equation which is plot the shape of the potential function}.
Then, we plot the shape function ${\cal U}_N\equiv{\cal V}_N(r,T)/N^2$ at different values of $M_g(T)/T$, to search the vacuum structure; the plots are displayed in Fig.~\ref{Fig:V-s}. For the $N=2$ case, there is only one minimum at $r\neq 1$, and eventually, only the minimum at $r=1$ survives as the increasing $M_g(T)/T$; see the left panel of Fig.~\ref{Fig:V-s}. So, in this case, the transition from the deconfined phase to the confining phase is crossover. On the contrary, for the case $N\geq 3$, ${\cal U}_N$ has two minimums when $M_g(T)/T$ approaches 2.7, with one located at $r=1$ and the other one at $r\neq 1$. It implies that the deconfinement phase transition is first order. Furthermore, via the degeneracy condition, we can determine the critical temperature $T_c$ in the unit of $M_g(T_c)$; see the right panel of Fig.~\ref{Fig:V-s}.

The phase transition behavior predicted by the model is consistent with the results of lattice simulation. Hence, it is of importance to understand what causes the qualitative difference between the shape functions for $N=2$ and $N\geq 3$. To that end, we investigate the shape function near $r=1$, and hence it is convenient to set $t=1-r$; then, we expand it around $t=0$,keeping the irrelevant terms up to ${\cal O}(t^5)$,
\begin{equation}\label{}
\begin{split}
{\cal U}_N=&t^2 \left(\frac{\pi ^2 {c_g} \csc ^2\frac{\pi
   }{N}}{N^2}+\frac{6}{N^4}-\frac{5}{N^2}-1\right)+t^3 \left(\frac{2 \pi ^3
  c_g\cot \frac{\pi    }{N} \csc ^2\frac{\pi
   }{N}}{N^3}-\frac{8}{N^4}+\frac{10}{N^2}-2\right)
   \\ &+t^4 \left[ c_g\frac{\pi ^4
    \csc ^2\frac{\pi}{N}}{3N^2}\L \f{3}{N^2}+\frac{9  \cot
   ^2\frac{\pi}{N}  }{N^2}-1\R+\frac{3}{N^4}-\frac{5}{N^2}+2\right],
\end{split}
\end{equation}
where $c_g\equiv (M_g/T)^2 K_2(M_g/T)$. The special property of $N=2$ case is that the cubic term vanishes and therefore there is no barrier. For $N>2$, the cubic term is present and moreover carries a positive coefficient (attributed to the gluon potential), and as a consequence ${\cal U}_N$ is able to give rise to the first order deconfinement phase transition.

We end up this section with a comment on the Haar-type model~\cite{Mocsy:2003qw,Fukushima:2003fw,Roessner:2006xn,Fukushima:2008wg}, whose potential is supposed to resemble  
\begin{equation}
    V=-a(T)l^2/2+b(T)\log H_N(L).
\end{equation}
The first term of this potential comes from the kinetic term of $SU(N)$ theory which also exists in our model. The second term, characterzied by $H_N(L)$, actually is known as the Vandermonde determinant interaction of the $SU(N)$ theory; it appears mathematically to define an integration over a continuous group, which requires an invariant Haar group measure. Such an interaction is non perturbative, and is argued to be consistent with the picture of ghost dominance. Explicitly, the integrand functional of our effective potential $-\Omega(M_g=0)$ in the infrared regime ($E,p\rightarrow 0$) resembles the Vandermonde determinant interaction. In other words, roughly speaking it is a part of the ghost contribution.

\section{Thermodynamics}

Although the model can surprisingly describe the order of deconfinement phase transition for any $N$, it is still important to check if it is able to correctly account for the thermodynamics above $T_c$, in particular in the semi-QGP region around $1.4T_c$ where the nonperturbative effect is significant. We have to rely on the temperature varying  $M_g(T)$, with $M_g(T_c)$ fixed (traded with the critical temperature $T_c$), to do this job. 

We should start from fitting latent heat. Because the quasi gluon mass is temperature dependent, the latent heat is sensitive to $dM_g/dT$ at $T_c$. Actually, the latent heat data can fix the value of this derivative at $T_c$, which is crucial to fit $M_g(T)$ via thermodynamics.

\subsection{Latent heat and determination of $dM_g(T)/dT$ at $T_c$}

From the thermodynamics, it is known that the latent heat $L_N$ released during the first order phase transition is the energy density difference between two vacua, $L_N= \varepsilon_d-\varepsilon_c$, with subscripts $d$ and $c$ denoting for the deconfined and confining vacuum, respectively. Then, using the second law of thermodynamics, one can find that
\begin{equation}
\begin{split}
    L_N
    =T_c\frac{\partial \Delta P}{\partial T}-\Delta P
  =-T_c\frac{\partial \Delta \mathcal{V}(T)}{\partial T}|_{T=T_c}+\Delta \mathcal{V}_N(T_c),
\end{split}
\end{equation}
where $\Delta \mathcal{V}_N=\mathcal{V}_{d}-\mathcal{V}_{c}$ is the potential energy difference, vanishing at $T_c$. As a consequence, the latent heat is determined by the entropy part, i.e., $ L_N= -T_c\frac{\partial \Delta \mathcal{V}(T)}{\partial T}|_{T=T_c}$.

Note that so far we can not guarantee the confining vacuum at $s=1$ is indeed the absolute minimum below $T_c$, but it is simply an inference of the requirement that latent heat should be positive: It means $\partial \mathcal{V}_d/\partial T_c< \partial\mathcal{V}_c/\partial T_c$, and moreover at $T_c$ two vacua is degenerate, $\mathcal{V}_d=\mathcal{V}_c$, so, below $T_c$ one indeed has  $\mathcal{V}_d<\mathcal{V}_c$. 

We are now in the position to calculate the latent heat in our model. The straightforward calculation of the temperature derivative of the effective potential gives
\begin{equation}
\begin{split}
    \frac{\partial\mathcal{V}(r,T)}{\partial T}
    &=-\frac{3M_g^2N^2}{2\pi^2}l_N^2\Bigg[ 4 T K_2(M_g/T)+K_1(M_g/T)\bigg(M_g-T\frac{dM_g}{dT}\bigg) \Bigg]
    \\
    &-N^2\frac{2\pi^2}{45}\Bigg[(-1+r)^2(-1-2r+2r^2)-\frac{5(-1+r)^2r^2}{N^2}+\frac{r^3(-4+3r)}{N^4} \Bigg]T^3,
\end{split}
\end{equation}
In our model, the confining vacuum is always located at $r=1$ or $l_N=0$, and thus the contribution of the above derivative in this vacuum is a trivial term. Then, the latent heat is determined by the contribution from the deconfined vacuum, 
\begin{equation}\label{latantH}
\begin{split}
    \frac{L_N}{(N^2-1)T_c^4}
    &=\frac{N^2}{N^2-1}\Bigg(\frac{3l_{N,d}^2}{2\pi^2}\bigg(\frac{M_g}{T_c}\bigg)^2\Bigg[ 4 K_2(M_g/T_c)+K_1(M_g/T_c)\bigg(\frac{M_g}{T_c}-\frac{dM_g}{dT_c}\bigg) \Bigg]
    \\
    &+\frac{2\pi^2}{45}\Bigg[(-1+r_d)^2(-1-2r_d+2s_d^2)-\frac{5(-1+r_d)^2r_d^2}{N^2}+\frac{r_d^3(-4+3r_d)+1}{N^4}\Bigg]\Bigg),
\end{split}
\end{equation}
where $r_d$ (or $l_{N,d}$) is the value of $r$ (or $l_N$) in the deconfined vacuum, numerically calculated by virtue of the tadpole Eq.~(\ref{tadpole}), shown in Table.~\ref{table-dMg}.

On the other hand, for $N=3,...8$, the current lattice data gives the following behavior of latent heat~\cite{Datta:2010sq} 
\begin{equation}\label{Lat}
\frac{L_N}{(N^2-1)T_c^4} \simeq 0.388-\frac{1.61}{N^2},
\end{equation}
We require the calculated latent heat Eq.~(\ref{latantH}) to fit it. For the given $N$, Eq.~(\ref{latantH}) just contains a single parameter, $dM_g(T)/dT$ at $T_c$, and therefore its value can be uniquely fixed. We show the results in Table.~\ref{table-dMg}. The resulting values typically are around $-10$ for all $N$, indicating a sharp increasing of quasi-gluon mass as the temperature drops down to $T_c$ from above. This is a well-understood behavior since it can be regarded as a sign of the ``strongest" nonperturbative effect near $T_c$. 
\begin{table} 
\begin{tabular}
{ |p{2.5cm}||p{1.5cm}|p{1.5cm}|p{1.5cm}|p{1.5cm}|p{1.5cm}|p{1.5cm}|p{1.5cm}|   } 
 \hline
Color number& $N=3$ & $N=4$ &$N=5$&$N=6$&$N=7$&$N=8$&$N\rightarrow\infty$\\
 \hline
$r_d$& 0.5605 &0.5186& 0.5073&0.5033&0.5016&0.5009&0.5004\\
 \hline
$l_d$& 0.5910 &0.6300& 0.6380&0.6398&0.6400&0.6396&0.6367\\
 \hline
$M_g(T_c)/T_c$& 2.7499 &2.7203& 2.7126&2.7099&2.7088&2.7083&2.7077\\ 
 \hline
$dM_g(T_c)/dT_c$& -5.7727 &-7.9951& -9.2891&-10.0965&-10.6261&-10.9954&-12.3376\\
 \hline
$L_N/(N^2-1)T_c^4$& 0.2091 &0.2874& 0.3236&0.3433&0.3551&0.3628&0.3880\\
 \hline
\end{tabular}
\caption{Effective mass, the position of deconfined vacuum and, latent heat around the critical temperature for different color number $N$. }  
\label{table-dMg}
\end{table}

\subsection{Fit $M_g$ with the thermal quantity using machine learning}

According to the original idea of QPM, the proper temperature dependence beyond $T_c$ of quasi-gluon mass is supposed to successfully explain the thermodynamics of the hot PYM system up to the high $T$ region. Here, the main thermodynamic observables of interest are the pressure $p$, the energy density $\epsilon$, and the entropy density $s$. Actually, they are not independent quantities. In particular, if one has $p$, then $\epsilon$ and $s$ can be calculated by the second law of thermodynamics
\begin{equation}
\begin{split}
    \epsilon=T\frac{dp}{dT}-p,\ \ \ \ s=\frac{\epsilon+p}{T}.
\end{split}
\end{equation}
The one loop calculation leads to $p=-{\cal V}_{eff}$, given in Eq.~(\ref{EffP}). Currently, their lattice data is available only for $N=3,4,6$~\cite{Datta:2010sq}. However, as stated in the introduction, the lattice data demonstrates $N$ scaling property, which means $P_M=\frac{M^2-1}{N^2-1}P_N$ and the latent Eq.~(\ref{Lat}), and thus we also have ``data" for other $N$ values by simple extrapolation, for instance to $N=5,8$ used later.

In the QPM, it is known that the SB limit can be trivially recovered. The most challenging range is the so-called semi-QGP region $T\in(T_c,3T_c)$, where the deviation to the blackbody behavior becomes more and more remarkable as $T$ approaching $T_c$. In the previous discussion, we have used effective potential  Eq.~(\ref{def:EP}), which is based on the high and low temperature expansion, to analyze the phase transition at $T_c$. Nevertheless, we do not have such a simple analytic expression to analyze thermadynamics. It is well expected that the low temperature expansion just holds very near $T_c$ and soon becomes not reliable in the higher temperature region. Hence, we should use its complete expression:
\begin{equation}\label{Edef:EP}
%\begin{aligned}
\mathcal{V}_{eff} 
\begin{aligned}[t]
&=\frac{T^4}{2\pi^2}\int_0^\infty dx x^2
\Bigg((N-1)\log[1-e^{-\hat{E}(x,M_g,T)}]
\\
&+\sum_{i=1}^{N}(N-i)\log[1+e^{-2\hat{E}(x,M_g,T)}-2e^{-\hat{E}(x,M_g,T)}\cos(2\pi\frac{i r}{N})]\Bigg)+...,
\end{aligned}
% \\
% &-\frac{\pi^2T^4}{90}+\frac{N^2\pi^2T^4}{90}\Bigg[ (s-1)^2(2s^2-2s-1)-\frac{5(s-1)^2s^2}{N^2}+\frac{s^3(3s-4)}{N^4} \Bigg]
%\end{aligned}
\end{equation}
where $\hat{E}(x,M_g,T)=\sqrt{x^2+(M_g/T)^2}$ and dots denote for the remaining term that does not need summation. 

\begin{figure}[htbp]
\centering 
\includegraphics[width=0.45\textwidth]{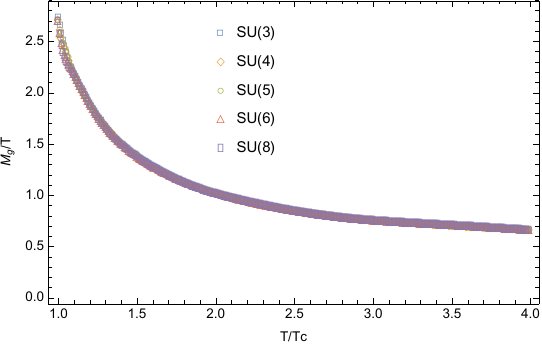} 
\includegraphics[width=0.45\textwidth]{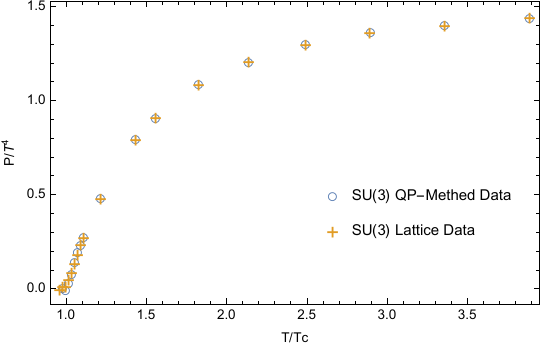} 
\includegraphics[width=0.45\textwidth]{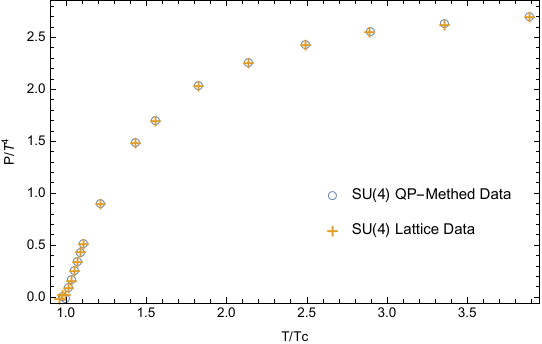} 
\includegraphics[width=0.45\textwidth]{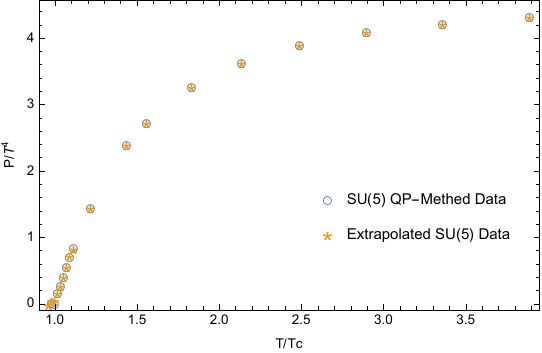} 
\includegraphics[width=0.45\textwidth]{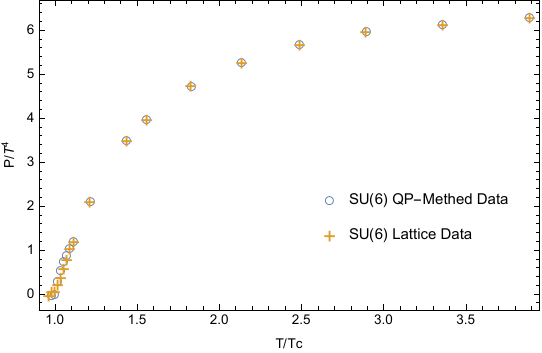} 
\includegraphics[width=0.45\textwidth]{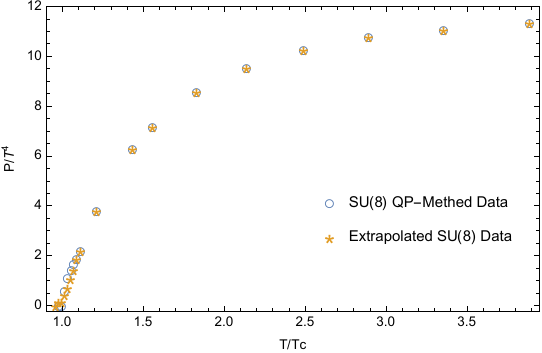} 
\caption{The first panel: The fitted $M_g/T$ as a function of temperature for $N=3,4,5,6,8$. The second to the sixth panels: Fitting $p/T^4$ in our model for various color number.} \label{Fig:fit}
\end{figure}

Then, we try to obtain the interpolation function of the fitted effective gluon mass $M_g(T)$ for $N=3,4,5,6,8$, through the method of machine learning. Physical Information Neural Network~\cite{Raissi:2019jcp} provides us with a flexible and accurate method for the fitting task. It treats functions of any complexity under fitting as a neural network, and the training goal is making the neural network satisfying the required partial differential relationships (such as partial differential equations and boundary conditions) and the given data points values. In our work, we use two separate deep neural networks $ M_g(T)$ and $ r(T)$ for the fitting task, and our training goal is making $ M_g(T)$ and $ r(T)$ to satisfy: 
\begin{itemize}
\item the extreme condition for the deconfined vacuum,
\begin{equation}
\begin{split}
\frac{\partial \mathcal{V}(r,T,N)}{\partial r}|_{r=r_d(T)}=0;
\end{split}
\end{equation}
\item the degeneracy between  the deconfined vacuum and the confining vacuum,
\begin{equation}
\begin{split}
\mathcal{V}(r,T,N)|_{r=r_d(T=T_c),T=T_c}=\mathcal{V}(1,T_c,N);
\end{split}
\end{equation}
\item mass parameter relationship in Table.~\ref{table-dMg} and 
\item  the lattice data for thermodynamics.
\end{itemize}
We implement the task using TensorFlow2.0~\cite{10.5555/3026877.3026899}, both $ M_g(T)$ and $ r(T)$ containing 7 hidden layers, each of which  includes 64, 128, 256, 512, 256, 128, 64 neutrons respectively. For the complexity of our problem, we should adopt a two-step training: We pretrain $ M_g(T)$ and $ r(T)$ to fit the lattice data first, and then fine adjust $ M_g(T)$ and $ r(T)$ to satisfy other fitting requirements. Such a procedure motivates us to divide the training samples into two types, the first type satisfies the lattice thermodynamic data at $T_c$, and the second type is 128 points randomly distributing in the temperature region $[T_c,4T_c]$, which meet the other three theoretical conditions listed above. For more details, please check the code in Github~\footnote{https://github.com/JGuoHep/QuasiParticle}. The fitted $M_g(T)$  is shown in the first panel of Fig.~\ref{Fig:fit}, and the perfect fitting of pressure above $T_c$ is displayed in other panels of Fig.~\ref{Fig:fit}.

\begin{figure}[htbp]
\centering 
\includegraphics[width=0.45\textwidth]{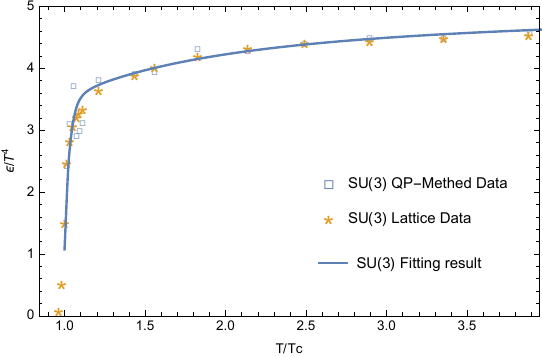} 
\includegraphics[width=0.45\textwidth]{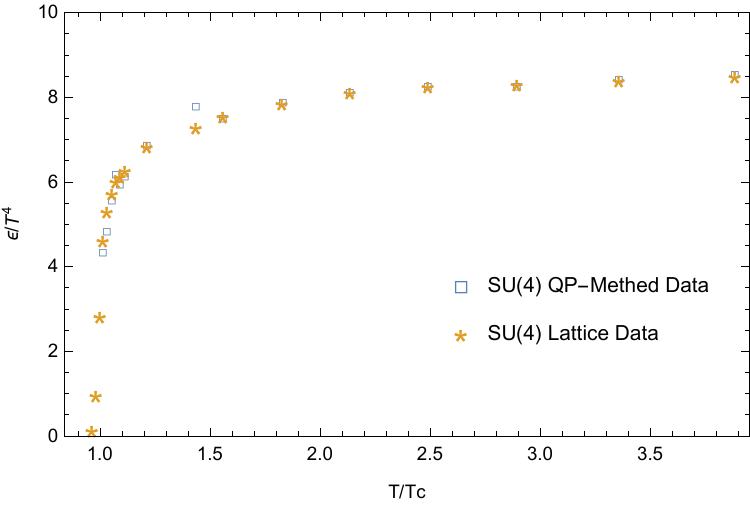} 
\includegraphics[width=0.45\textwidth]{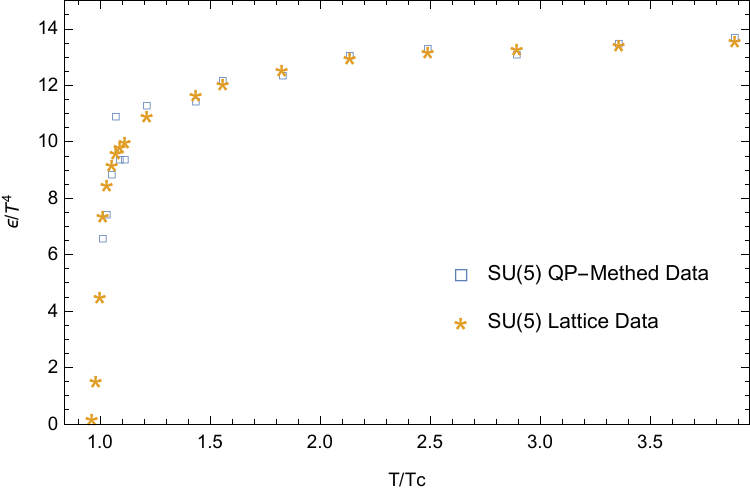} 
\includegraphics[width=0.45\textwidth]{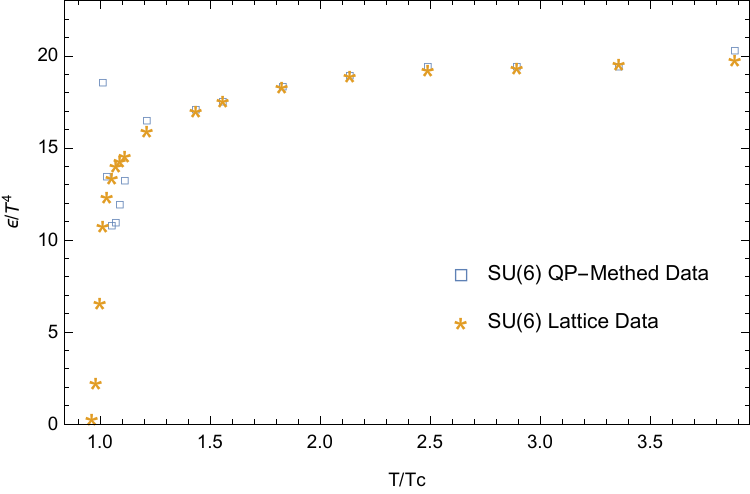} 
\includegraphics[width=0.45\textwidth]{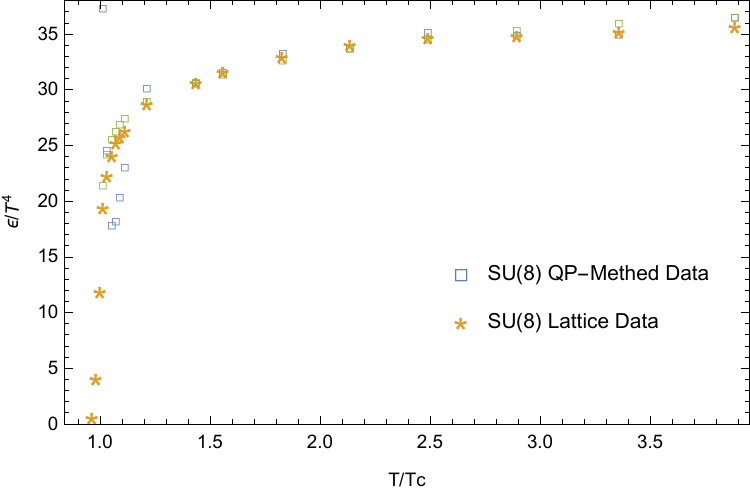} 
\includegraphics[width=0.45\textwidth]{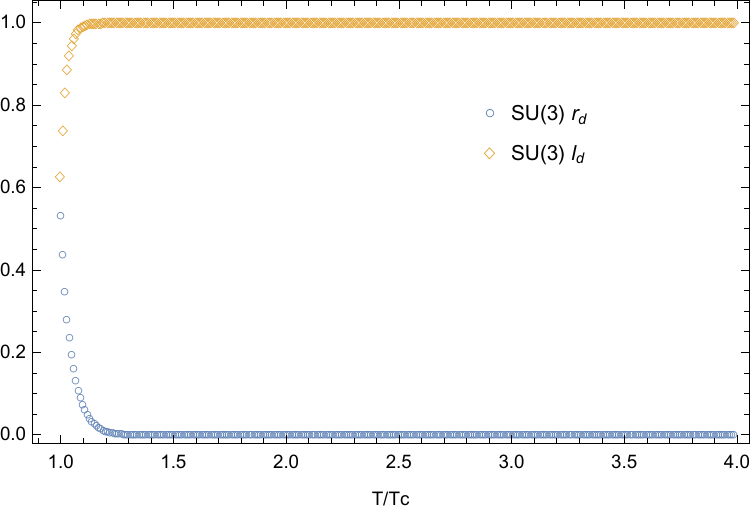} 
\caption{ From the first to the fifth panels: energy density for $N=3,4,5,6,8$ (the blue square denoting for model prediction using the numerical function $M_g$ from machine learning; the yellow star denoting for lattice data; in the first panel, for comparison, the prediction using the smooth fitting function Eq.~(\ref{lmgfit}) for $M_g$ labeled as the blue line). The last panel: the value of the order parameter in the deconfinement phase  $l_d$ (or $r_d$) varying with temperature for $N=3$. } \label{Fig:E-SLT}
\end{figure}

With the fitted $M_g(T)$, one can plot the energy density $\epsilon$, shown in Fig.~\ref{Fig:E-SLT}. From the first five plots one can see that our model predictions fairly well match the lattice data for all $N$, except that the point around $1.3T_c$ always mildly deviates from the lattice result. The reason is that our training did not include energy density data, and the resulting numerical function $M_g(T)$ is continuous but its derivative is discontinuous (retraining may lead to slight improvement). However, if we instead use the smooth fitting function Eq.~(\ref{lmgfit}) obtained later rather than the original numerical function, the calculated energy density can fit well with the lattice data, as shown in the example of $SU (3)$ in Fig.~\ref{Fig:E-SLT}.

We also plot the value of the order parameter in the deconfinement phase, the Polyakov loop $l_d$ or equivalently $r_d$ here. We only show the $SU(3)$ case in the last panel of Fig.~\ref{Fig:E-SLT},  which has been studied on the lattice; from the plot we can see that as the temperature rises, the value of $l_d/r_d$ soon approaches 1/0. The overall trend is right, but the $l_d(T)$ predicted in our model reaches 1 faster than the lattice result. This issue might be resolved by considering the dressing propagators, which introduce more parameters; for comparison, here we have only one parameter, $M_g$. We leave this study to the future work.

%our way of training which does not include data of energy density, 

%rendering that the numerical function $M_g(T)$ is a continuous function, its derivative is not continuous

%although  fitted by machine learning is a continuous function, its derivative is not continuous. However, although the values of some points deviate, 

The $M_g(T)$ is supposed to depend on $N$: Although $M_g(T_c)/T_c$ is almost universal determined by the condition of degeneracy, $dM_g(T)/dT$ takes different values at $T_c$ for different color number for the sake of correct latent heat, see Table.~\ref{table-dMg}. However, it is found that the fitted $M_g(T)$ are almost the same, which leads us to conjecture that this is an universal behavior for all $N$~\footnote{For large $N$ this is trivial, because the $N$ dependence of the observables in our model is scaled out, well consistent with the lattice data. But it is not trivial that it is true also for $N=3,4$. By contrast, in the polynomial model~\cite{Kang:2021epo}, the fitting parameters in the small $N$ cases are very different than those in the large $N$ cases. We guess it is attributed to the exponential dependence of the fitting parameter $M_g(T)$.}. By the way, one can check the invalidation of low temperature expansion in the region $T\gtrsim 1.4T_c$: The ratio $M_g/T$ drops to $\approx 1.5$ as $T$ increases to $1.4T_c$, and then from Eq.~(\ref{LTE}) one can see that the next leading order is only suppressed by a factor $K_2(2\times 1.5)/K_2(1.5)\sim 0.1$. 

\begin{table}[htbp]
\begin{tabular}{|c|c|c|c|}
%{ |p{1.5cm}|p{1.5cm}|p{1.5cm}||p{1.5cm}|   } 
 \hline
$\alpha$ & $\beta$ & $\gamma$ & RMSD \\
 \hline
 0.029534 & 1.130884 & 1.541299 & 0.015707\\ 
 \hline
 0 & 1.186505 & 1.570699 & 0.016224\\
 \hline
 \hline
 void & $\lambda$ & $T_s$ & RMSD\\
 \hline
 void & 10.843298 & -8.336149 & 0.081482\\
 \hline
\end{tabular}
\caption{$\alpha$, $\beta$ and $\gamma$ are fitting parameters in the qausigluon mass ansatz Eq.(\ref{lmgfit}), while $\lambda$ and $T_s$ are fitting parameters in the conventional ansatz Eq.(\ref{QPM:mass}). }
\label{table-mgfit}
\end{table}

\begin{figure}[htbp]
\centering 
\includegraphics[width=0.45\textwidth]{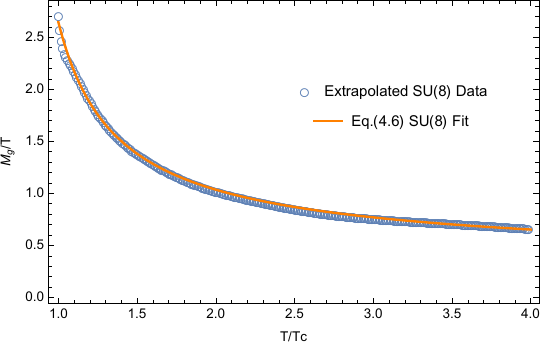} 
\includegraphics[width=0.45\textwidth]{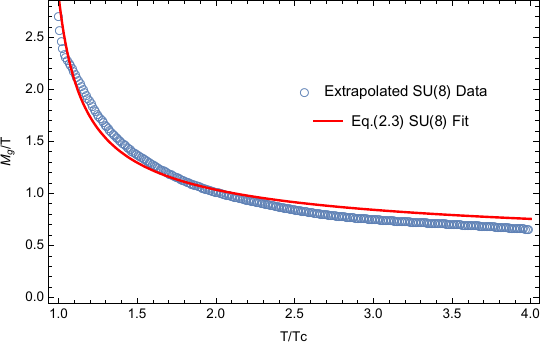} 
\caption{The fitting for the quasigluon mass ansatz Eq.({\ref{lmgfit}}) (left panel) and Eq.~(\ref{QPM:mass}) (right panel), where the fitting paramers are given in Table.~\ref{table-mgfit}.} \label{Fig:fitMg}
\end{figure}

Actually, the $N$ universal behavior of quasigluon mass is encoded in the quasigluon mass in the HTLpt; see the formula Eq.~(\ref{QPM:mass}) where $N$ cancels. At this point, our model is consistent with the HTLpt effective mass. So, it is anticipated that the interpolation function can be fitted by the $M_g(T)$ with the function given in  Eq.(\ref{QPM:mass}), with two parameters $\lambda$ and $T_s$. We also try another function with three parameters
\begin{equation}\label{lmgfit}
 M_g(T)=\alpha T+\beta T/\log(\gamma T/T_c).
\end{equation}
which is recently adopted in Ref.~\cite{Islam:2021qwh}. Note that unlike the conventional $M_g(T)$ ansatz, which simply goes to the HTLpt quasigluon mass in the high $T$ region, Eq.~(\ref{lmgfit}) does not. The fitted parameters for both functions of $M_g(T)$ are shown in Table.~\ref{table-mgfit}. The latter has better quality, which can be seen from the comparison in two panels of Fig.~\ref{Fig:fitMg}. This may raise the issue of well consistence between our model with the HTLpt in the higher $T$ region, and we will come back to this point in the Section of conclusion and discussion. Besides, for the function Eq.~(\ref{lmgfit}), from Table.~\ref{table-mgfit} one can see that the values of the $\alpha$ parameter are far smaller than the other two parameters, which means that it is almost irrelevant to fitting. So, we tried the fitting with the vanishing $\alpha$, to find that it works equally well.

%and From the Fig.~\ref{Fig:fitMg} and Table.~\ref{table-mgfit}, one can see that the Fitting result by using Eq.~(\ref{QPM:mass}) is not good as one from Eq.~(\ref{lmgfit}). This is the reason of why we introduce a different function from to do the fitting rather than directly using the traditional quasi-particle mass. 

%Fig.~\ref{Fig:fitMg} and We tried two forms of $M_g(T)$ to fit the  interpolation function. The first one is Eq.(\ref{QPM:mass}); The second function is inspired by the quasigluon mass in the HTLpt and have the following form
%we adopt the following function of quasi gluon mass to fit the obtained interpolation function

%Then, the universal quasigluon mass conjecture might be ``tested" by the extrapolation of the data. It is found that the prediction is well supported by the extrapolated data for $N=5,8$, shown in First panel of Fig.~\ref{Fig:fitMg}.

%The fitting result of thermal lattice data with our model is also showed in 5-8 panel of Fig.\ref{Fig:fit}. We discussed the case of color numbers $N=3,4,5,6,8$. The lattice data which we used\cite{Datta:2010sq} only have the result for $N=3,4,6$. However, those data show $N$ scaling property which means $P_M=\frac{M^2-1}{N^2-1}P_N$. So, we generate those data into other color number $N=5,8$. As showed in the plot, our model can fit the lattice data very well.

\section{Conclusion and discussion}

The HTL resummation in the quasi-particle picture reveals that QGP is a pool of weakly interacting quasigluons for $T\gtrsim 2T_c$. Such a picture is further used in the QPM to describe QCD thermodynamics down to $T_c$ and works fairly well. The crucial idea is that the quasigluon mass could ``absorb" strong interaction and merely leaves weak interactions on quasigluons. In this work we attempt to embed this idea to the massive PYM~\cite{Reinosa:2014ooa}, introducing a temperature-dependent quasigluon mass in the effective $SU(N)$ PYM Lagaragian Eq.~(\ref{def:la}). Via the standard perturbative calculation, we obtain an effective model that can successfully explain the critical behavior for any $N$, not also the first order deconfinement phase transition for $N>2$ but also the crossover for $N=2$. Moreover, the lattice data of thermodynamics can be fitted via the single parameter $M_g(T)$, which is found to demonstrate the $N$-universal behavior, based on the available case $N=3,4,6$. This is supported by the HTLpt quasigluon mass, but now is extended to the semi-QGP region, and might convey some secrets of the non-perturbative effects. We look forward to the future lattice data for other $N$, in particular, $N=5,8$ whose ``lattice data" is obtained by extrapolation via the $N$-scaling law,  to test the universal quasigluon mass conjecture.

Fitting $M_g(T)$ via a function that well matches with the HTLpt quasigluon mass does not have a very good quality, and it may be improved by considering the dressing propagator of the gluons~\cite{Braun:2007bx}. Then, the modified model contains more parameter and have the potential to deal with more detailed problems.
%and moreover of the hot $SU(N)$ PYM system.

We are capable of conducting a unified analysis of all $N$, depending on the assumption of uniform eigenvalue distribution of the temporal background, which reduces the effective potential to the one-dimensional case. But it is based on the eigenvalue repulsion and a more solid argument may be necessary.

\noindent {\bf{Acknowledgements}}

This work is supported in part by the National Key Research 
 and Development Program of China Grant No. 2020YFC2201504 and in part by the National Science Foundation of China (11775086).

\noindent {\bf{Note added}}

Right before the submission of this work, the work by Fu-Peng Li etc.~\cite{Li:2022ozl} appeared on arxiv. They also utilize the machine learning to reconstruct QCD equation of state in the QPM picture, which may have partial overlap with our work.

\appendix

\section{Derivation of the generating function in the Landau-DeWitt gauge }\label{GFcomputation}

%\subsection{B}\label{GFcomputation}

The complete Faddeev-Popov Lagrangian in the Landau-DeWitt gauge reads 
\begin{equation}
\mathcal{L}=-\frac{1}{2g^2}{\rm tr}(F_{\mu\nu}F^{\mu\nu})+\bar{D}_\mu \bar{c}^a D^\mu c^a+ih^a \bar{D}_\mu\hat{A}^{\mu,a}+\f{1}{2}M_g(T)A^a_{tr,\mu} A^{a,\mu}_{tr},
\end{equation}
We are considering the constant background $\bar{A}^a_\mu=\bar{A}^a_0\delta_{\mu 0}$ and keep only the quadratic terms. 
% relevant to our discussions are collected in
% \begin{equation}
% \mathcal{L}=\frac{1}{2}\hat{A}_\mu^a(D^{-1})_{ab}\hat{A}^{\mu,b}+ih^a \bar{D}_\mu\hat{A}^{\mu,a}+\bar{D}_\mu \bar{c}^a \bar{D}^\mu c^a,
% \end{equation}
% where the first term has been explicitly . 
Then the action can be split into two parts $S_{A,h}$ and $S_c$, where
\begin{equation}
\begin{split}
    & S_{A,h}=\int d^4x\left[ \frac{1}{2}\hat{A}_\mu^a(D^{-1})_{ab}\hat{A}^{\mu,b}+ih^a \bar{D}_\mu\hat{A}^{\mu,a}\right]
    \\
    & S_c=\int d^4x  [\bar{D}_\mu \bar{c}^a \bar{D}^\mu c^a],
\end{split}
\end{equation}
where $D^{-1}$ is given in Eq.~(\ref{D-1}). 

Now we come to deal with the first part of the action. We can rewrite this action in its color diagonalization basis $\tilde{A}^a_\mu$ and $\tilde{h}^a$, to get 
\begin{equation}
\begin{split}
    S_{A,h}
    &=\int d^4x\left[ \frac{1}{2}\tilde{A}_\mu^a(\tilde{D}^{-1})_{a}\tilde{A}^{\mu,a}+i\tilde{h}^a \tilde{D}_\mu^a\tilde{A}^{\mu,a}\right]
    \\
    &=\int\frac{d^4p}{(2\pi)^4}\left[ \frac{1}{2}\tilde{A}_\mu^a(\tilde{D}^{-1})_{a}\tilde{A}^{\mu,a}+i\tilde{h}^a \tilde{D}_\mu^a\tilde{A}^{\mu,a}\right],
\end{split}
\end{equation}
where $(\tilde{D}^{-1})_{a}$ and $\tilde{D}_\mu^a$ in the momentum space are respectively given by
\begin{equation}
\begin{split}
    &\tilde{D}^{-1}_{a}(p)\equiv D(M_g)_a^{-1}=(p_0-A_a)^2+|\vec{p}|^2+M_g^2,
    \\
    &\tilde{D}_\mu^a\tilde{D}^{\mu,a}\equiv D(0)_a^{-1}=(p_0-A_a)^2+|\vec{p}|^2.
\end{split}
\end{equation}

To integrate this action through path integration, we must do the quadratic partition between $\tilde{A}^a$ and $\tilde{h}^a$. After a tedious quadratic partition, the action takes the form of
\begin{equation}
\begin{aligned}
    S_{A,h}=\int\frac{d^4p}{(2\pi)^4}\frac{1}{2}
    \left[
    (\tilde{D}^{-1})_{a}(\tilde{A}_\mu^a+i\frac{\tilde{D}^a_\mu}{\tilde{D}^{-1}_a}\tilde{h}^a)(\tilde{A}^{\mu,a}
    +i\frac{\tilde{D}^{\mu,a}}{\tilde{D}^{-1}_a}\tilde{h}^a)+\frac{\tilde{D}_\mu^a\tilde{D}^{\mu,a}}{\tilde{D}^{-1}_a}\tilde{h}^a \tilde{h}^a
    \right].
\end{aligned}
\end{equation}
Now define a new field $\mathcal{A}^a_\mu=\tilde{A}_\mu^a+i\frac{\tilde{D}^a_\mu}{\tilde{D}^{-1}_a}\tilde{h}^a$, and one can rewrite the original mixed action as 
\begin{equation}
    S_{A,h}=\int\frac{d^4p}{(2\pi)^4}\frac{1}{2}
    \left[
    \mathcal{A}^a_\mu D(M_g)^{-1}_{a}\mathcal{A}^{\mu,a}+\frac{D(0)^{-1}_a}{D(M_g)^{-1}_{a}}\tilde{h}^a\tilde{h}^a
    \right].
\end{equation}
The redefined Nakanishi-Lautrup field $\tilde{h}^a$ now gains a mass, and its propagator is a combination of the masssive and massless propagators, which is a result of the Landau-Dewitt gauge. Then the 1-loop effective action is given by
\begin{equation}\label{ZAh}
\begin{split}
    \log{Z}_{A,h}
    &=\frac{1}{2}\log\det \left[ \frac{\delta^2 S_{h,A}}{\delta\phi_i^a\phi_j^a}\right]
    \\
    &=2\log\det(D(M_g)^{-1})-\frac{1}{2}\log\det(D(M_g)^{-1})+\frac{1}{2}\log\det(D(0)^{-1})
    \\
    &=\frac{3}{2}\log\det(D(M_g)^{-1})+\frac{1}{2}\log\det(D(0)^{-1}).
\end{split}
\end{equation}
where $\phi^a_i$ represent $\{\mathcal{A}^a_\mu,\tilde{h}^a\}$. Note that there is a overall factor 4 for the $\mathcal{A}^a_\mu$ contribution, denoting for four massive modes. But the Nakanishi-Lautrup field cancels one massive mode and effectively just leaves one massless mode.

The massless ghost contribution, taking into account its statistics, is simply given by
\begin{equation}
    \log{Z}_c=-\log\det(D(0)^{-1}).
\end{equation}
Its contribution is halved due to the massless mode of the Nakanishi-Lautrup field. Finally, the total effective action is
\begin{equation}
\begin{split}
    \log{Z}=\frac{3}{2}\log\det(D(M_g)^{-1})-\frac{1}{2}\log\det(D(0)^{-1}).
\end{split}
\end{equation}

\section{Calculating the pure gluonic generating function: the $SU(4)$ sample}\label{SU(4)}

In this appendix we present the details of calculating the pure glunoic part, i.e., the first term of the second line of Eq.~(\ref{ZAh}), specified to $SU(4)$. Its Cartan generators are 
% Let us quote the the quadratic Lagrangian of the fluctuation field $\hat{A}_\mu$ below
% \begin{equation}
% \mathcal{L}^{(2)}=-\frac{1}{2}\hat{A}_{\mu}^a[\delta_{ab}g^{\mu\nu}\partial^2-f_{abc}(\partial^\nu\bar{A}^{\mu,c}+2g^{\mu\nu}\bar{A}^c_\rho\partial^\rho)+f_{acd}f_{cbe}g^{\mu\nu}\bar{A}^d_\rho\bar{A}^{\rho,e}+2f_{abc}\bar{F}^{\mu\nu,c}]\hat{A}^b_\nu.
% \end{equation}
% The independent field related to the confinement problem only contains the diagonal term which for $SU(4)$ is given by $\bar{A}_\mu=\sum_{i=3,8,15}\bar{A}^i_\mu T^i$ with the generators
\begin{equation}
T^3=\frac{1}{2}
\left(
\begin{matrix}
1&0&0&0\\
0&-1&0&0\\
0&0&0&0\\
0&0&0&0
\end{matrix}
\right),\ \ 
T^8=\frac{1}{2\sqrt{3}}
\left(
\begin{matrix}
1&0&0&0\\
0&1&0&0\\
0&0&-2&0\\
0&0&0&0
\end{matrix}
\right),\ \ 
T^{15}=\frac{1}{2\sqrt{6}}
\left(
\begin{matrix}
1&0&0&0\\
0&1&0&0\\
0&0&1&0\\
0&0&0&-3
\end{matrix}
\right).
\end{equation}
Now the propagators take the form of (the quasigluon mass can be trivially included)
% Suppose the background field is a uniform field and only contains time component $\bar{A}_\mu^i=\bar{A}^i_0\delta_{0\mu}$. Then the Lagrangian becomes
% \begin{equation}
% \mathcal{L}^{(2)}=-\frac{1}{2}\hat{A}_\mu^a(D^{-1})_{ab}\hat{A}^{\mu,b}
% \end{equation}
% where
\begin{equation}
(D^{-1})_{ab}=\delta_{ab}p^2+2i\sum_{i=3,8,15}f_{abi}\bar{A}^i_0 p_0-\sum_{i,j=3,8,15}f_{aci}f_{cbj}\bar{A}^i_0\bar{A}^j_0.
\end{equation}
After a careful calculation, one can get all the non-zero propagators 
\begin{equation}
\begin{split}
&(D^{-1})_{1,1}=p^2+(\bar{A}^3_0)^2\ \ (D^{-1})_{1,2}=2ip_0\bar{A}^3_0
\\
&(D^{-1})_{2,2}=p^2+(\bar{A}^3_0)^2\ \ (D^{-1})_{1,2}=-2ip_0\bar{A}^3_0
\\
&(D^{-1})_{3,3}=p^2
\\
&(D^{-1})_{4,4}=p^2+\frac{1}{4}(\bar{A}^3_0)^2+\frac{4}{3}(\bar{A}^8_0)^2+\frac{\sqrt{3}}{2}\bar{A}^{3}_0\bar{A}^{8}_0\ \ (D^{-1})_{4,5}=ip_0\bar{A}^3_0+\sqrt{3}ip_0\bar{A}^8_0
\\
&(D^{-1})_{5,5}=p^2+\frac{1}{4}(\bar{A}^3_0)^2+\frac{4}{3}(\bar{A}^8_0)^2+\frac{\sqrt{3}}{2}\bar{A}^{3}_0\bar{A}^{8}_0\ \ (D^{-1})_{5,4}=-ip_0\bar{A}^3_0-\sqrt{3}ip_0\bar{A}^8_0
\\
&(D^{-1})_{6,6}=p^2+\frac{1}{4}(\bar{A}^3_0)^2+\frac{4}{3}(\bar{A}^8_0)^2-\frac{\sqrt{3}}{2}\bar{A}^{3}_0\bar{A}^{8}_0\ \ (D^{-1})_{6,7}=-ip_0\bar{A}^3_0+\sqrt{3}ip_0\bar{A}^8_0
\\
&(D^{-1})_{7,7}=p^2+\frac{1}{4}(\bar{A}^3_0)^2+\frac{4}{3}(\bar{A}^8_0)^2-\frac{\sqrt{3}}{2}\bar{A}^{3}_0\bar{A}^{8}_0\ \ (D^{-1})_{7,6}=ip_0\bar{A}^3_0-\sqrt{3}ip_0\bar{A}^8_0
\\
&(D^{-1})_{8,8}=p^2
\\
&(D^{-1})_{9,9}=p^2+\frac{1}{4}(\bar{A}^3_0)^2+\frac{1}{12}(\bar{A}^8_0)^2+\frac{1}{12}(\bar{A}^{15}_0)^2+\frac{1}{2\sqrt{3}}\bar{A}^{3}_0\bar{A}^{8}_0+\frac{1}{2\sqrt{3}}\bar{A}^{3}_0\bar{A}^{15}_0+\frac{1}{6}\bar{A}^{8}_0\bar{A}^{15}_0
\\
&(D^{-1})_{9,10}=ip_0\bar{A}_0^{3}+\frac{i}{\sqrt{3}}p_0+\bar{A}_0^{8}+\frac{i}{\sqrt{3}}p_0\bar{A}_0^{15}
\\
&(D^{-1})_{10,10}=p^2+\frac{1}{4}(\bar{A}^3_0)^2+\frac{1}{12}(\bar{A}^8_0)^2+\frac{1}{12}(\bar{A}^{15}_0)^2+\frac{1}{2\sqrt{3}}\bar{A}^{3}_0\bar{A}^{8}_0+\frac{1}{2\sqrt{3}}\bar{A}^{3}_0\bar{A}^{15}_0+\frac{1}{6}\bar{A}^{8}_0\bar{A}^{15}_0
\\
&(D^{-1})_{10,9}=-ip_0\bar{A}_0^{3}-\frac{i}{\sqrt{3}}p_0+\bar{A}_0^{8}-\frac{i}{\sqrt{3}}p_0\bar{A}_0^{15}
\end{split}
\end{equation}

\begin{equation}
\begin{split}
&(D^{-1})_{11,11}=p^2+\frac{1}{4}(\bar{A}^3_0)^2+\frac{1}{12}(\bar{A}^8_0)^2+\frac{1}{12}(\bar{A}^{15}_0)^2-\frac{1}{2\sqrt{3}}\bar{A}^{3}_0\bar{A}^{8}_0-\frac{1}{2\sqrt{3}}\bar{A}^{3}_0\bar{A}^{15}_0+\frac{1}{6}\bar{A}^{8}_0\bar{A}^{15}_0
\\
&(D^{-1})_{11,12}=-ip_0\bar{A}_0^{3}+\frac{i}{\sqrt{3}}p_0+\bar{A}_0^{8}+\frac{i}{\sqrt{3}}p_0\bar{A}_0^{15}
\\
&(D^{-1})_{12,12}=p^2+\frac{1}{4}(\bar{A}^3_0)^2+\frac{1}{12}(\bar{A}^8_0)^2+\frac{1}{12}(\bar{A}^{15}_0)^2-\frac{1}{2\sqrt{3}}\bar{A}^{3}_0\bar{A}^{8}_0-\frac{1}{2\sqrt{3}}\bar{A}^{3}_0\bar{A}^{15}_0+\frac{1}{6}\bar{A}^{8}_0\bar{A}^{15}_0
\\
&(D^{-1})_{12,11}=ip_0\bar{A}_0^{3}-\frac{i}{\sqrt{3}}p_0+\bar{A}_0^{8}-\frac{i}{\sqrt{3}}p_0\bar{A}_0^{15}
\\
&(D^{-1})_{13,13}=p^2+\frac{1}{12}(\bar{A}^8_0)^2+\frac{1}{12}(\bar{A}^{15}_0)^2-\frac{1}{6}\bar{A}^{8}_0\bar{A}^{15}_0\ \ (D^{-1})_{13,14}=-ip_0\bar{A}^8_0+\frac{i}{\sqrt{3}}p_0\bar{A}^{15}_0
\\
&(D^{-1})_{14,14}=p^2+\frac{1}{12}(\bar{A}^8_0)^2+\frac{1}{12}(\bar{A}^{15}_0)^2-\frac{1}{6}\bar{A}^{8}_0\bar{A}^{15}_0\ \ (D^{-1})_{14,13}=ip_0\bar{A}^8_0-\frac{i}{\sqrt{3}}p_0\bar{A}^{15}_0
\\
&(D^{-1})_{15,15}=p^2
\end{split}
\end{equation}
It is observed that the $15\times 15$ propagator matrix in the color space $D^{-1}_{a,b}$ is a block diagonal matrix, consisting of three diagonal elements $p^2$ corresponding to the Cartan part and six $2\times 2$ submatrices corresponding to the non-Cartan parts. Concretely, these six matrices are
\begin{equation}
M_i=
\begin{bmatrix}
 p_0^2+\wt A_i^2-|\vec{p}|^2 & ip_0 \wt A_i\\
-ip_0 \wt A_i & p_0^2+\wt A_i^2-|\vec{p}|^2
\end{bmatrix},
\end{equation}
where $\wt A_i$ is a combination of $A_0$. From this expression we can see that the eigenvalues of $M_i$ must be $(p_0+A_i)^2-|\vec{p}|^2$ and $(p_0+A_i)^2+|\vec{p}|^2$. After a unitary diagonalization we can get $\wt D^{-1}\equiv U^\dagger D^{-1} U$ as
\begin{equation}
\begin{split}
&(\tilde{D}^{-1})_{1,1}=(p_0-\bar{A}_0^{3})^2-|\vec{p}|^2,\ \ 
(\tilde{D}^{-1})_{2,2}=(p_0+\bar{A}_0^{3})^2-|\vec{p}|^2
\\
&(\tilde{D}^{-1})_{3,3}=(\tilde{D}^{-1})_{8,8}=(\tilde{D}^{-1})_{15,15}=p^2
\\
&(\tilde{D}^{-1})_{4,4}=[p_0-\frac{1}{2}(\bar{A}_0^{3}+\sqrt{3}\bar{A}_0^{8})]^2-|\vec{p}|^2,\ \ 
(\tilde{D}^{-1})_{5,5}=[p_0+\frac{1}{2}(\bar{A}_0^{3}+\sqrt{3}\bar{A}_0^{8})]^2-|\vec{p}|^2
\\
&(\tilde{D}^{-1})_{6,6}=[p_0+\frac{1}{2}(\bar{A}_0^{3}-\sqrt{3}\bar{A}_0^{8})]^2-|\vec{p}|^2,\ \ 
(\tilde{D}^{-1})_{7,7}=[p_0-\frac{1}{2}(\bar{A}_0^{3}-\sqrt{3}\bar{A}_0^{8})]^2-|\vec{p}|^2
\\
&(\tilde{D}^{-1})_{9,9}=[p_0-\frac{1}{2}(\bar{A}_0^{3}+\frac{1}{\sqrt{3}}\bar{A}_0^{8}+\frac{1}{\sqrt{3}}\bar{A}_0^{15})]^2-|\vec{p}|^2
\\ 
&(\tilde{D}^{-1})_{10,10}=[p_0+\frac{1}{2}(\bar{A}_0^{3}+\frac{1}{\sqrt{3}}\bar{A}_0^{8}+\frac{1}{\sqrt{3}}\bar{A}_0^{15})]^2-|\vec{p}|^2
\\
&(\tilde{D}^{-1})_{11,11}=[p_0+\frac{1}{2}(\bar{A}_0^{3}-\frac{1}{\sqrt{3}}\bar{A}_0^{8}-\frac{1}{\sqrt{3}}\bar{A}_0^{15})]^2-|\vec{p}|^2
\\
&(\tilde{D}^{-1})_{12,12}=[p_0-\frac{1}{2}(\bar{A}_0^{3}-\frac{1}{\sqrt{3}}\bar{A}_0^{8}-\frac{1}{\sqrt{3}}\bar{A}_0^{15})]^2-|\vec{p}|^2
\\
&(\tilde{D}^{-1})_{13,13}=(p_0-\frac{1}{2}\frac{\bar{A}_0^{8}-\bar{A}_0^{15}}{\sqrt{3}})^2-|\vec{p}|^2,\ \ 
(\tilde{D}^{-1})_{14,14}=(p_0+\frac{1}{2}\frac{\bar{A}_0^{8}-\bar{A}_0^{15}}{\sqrt{3}})^2-|\vec{p}|^2.
\end{split}
\end{equation}
This leads to the quadratic Lagrangian written as
\begin{equation}
    \mathcal{L}=-\frac{1}{2}\wt A_\mu^a(\wt D^{-1})_{a}\wt A^{\mu,a},
\end{equation}
with $\wt D^{-1}={\rm diag}((p_0+A_1)^2-|\vec{p}|^2,(p_0+A_2)^2-|\vec{p}|^2,...,(p_0+A_{15})^2-|\vec{p}|^2)={\rm diag}\{ (p_0+A_a)^2-|\vec{p}|^2 \}$ where $A_a$ is zero or opposite numbers appearing in pairs. This structure is insured by the structural constant $f_{abc}$. One can check this structure for other $SU(N)$ theory. For example in $SU(3)$, the diagonal propagator is the same as the first eight propagators of $SU(4)$.

\section{Summation over the thermal modes}\label{GFcomputation1}

%\subsection{A}

In this appendix we explicitly implement the summation over the thermal modes present in Eq.~(\ref{Eac}), rewritten as
\begin{equation}\label{Z1}
\log Z=2 V{\rm tr_c}\int\frac{d^3\vec{p}}{(2\pi)^2}\nu(E_g),
\end{equation}
where we have introduced the function 
\begin{equation}\label{sumlog1}
\begin{split}
\nu(E_g)\equiv \sum_{n=-\infty}^\infty\log[\tilde{D}^{-1}_{aa}]
=\sum_{n=-\infty}^\infty\log[(\omega_n-A_a)^2+E_g^2],
\end{split}
\end{equation}
with $E_g^2=|\vec{p}|^2+M_g^2$. 
% \begin{equation}
% \log Z=2 V{\rm tr_c}\int\frac{d^3\vec{p}}{(2\pi)^2}\sum_{n=-\infty}^\infty\log[\tilde{D}^{-1}_{aa}(\omega_n,|\vec{p}|)].
% \end{equation}
% First, we define the function $\nu$ as 
% \begin{equation}\label{sumlog1}
% \begin{split}
% \nu(E_g)=\sum_{n=-\infty}^\infty\log[\tilde{D}^{-1}_{aa}]
% =\sum_{n=-\infty}^\infty\log[(\omega_n-A_a)^2+E_g^2].
% \end{split}
% \end{equation}
% where $E_g^2=|\vec{p}|^2+M_g^2$ and the generating function becomes
% \begin{equation}
% \log Z=2 V{\rm tr_c}\int\frac{d^3\vec{p}}{(2\pi)^2}\nu(E_g),
% \end{equation}
To pull out the object to be summed from the logarithm, we differentiate $\nu(E_g)$ with respect to $E_g$, 
\begin{equation}\label{}
\begin{split}
\frac{\partial\nu(E_g)}{\partial E_g}
&=\sum_{n=-\infty}^\infty\frac{2E_g}{(\omega_n-A_a)^2+E_g^2}
=\frac{1}{\pi T}\sum_{n=-\infty}^\infty\frac{E_g/2\pi T}{(n-A_a/2\pi T)^2+(E_g/2\pi T)^2}.
\end{split}
\end{equation}
Such a series can be summed explicitly, to get
\begin{equation}\label{sumlog2}
\begin{split}
\frac{\partial\nu(E_g)}{\partial E_g}
=\frac{1}{T}\frac{\sinh^2\L E_g/2\pi T\R}{\sin^2(A_a/2\pi T)+\sinh^2(E_g/2\pi T)}\left( 1+\frac{e^{-E_g/ T}}{1-e^{-E_g/ T}} \right).
\end{split}
\end{equation}
Then, integrating both sides over $E_g$, we have
\begin{equation}\label{sumlog3}
\nu(E_g)=\frac{E_g}{T}+2\log \left[ \sqrt{\L1-e^{-i\frac{A_a}{T}}e^{\frac{-E_g}{T}}\R\L1-e^{i\frac{A_a}{T}}e^{\frac{-E_g}{T}}\R} \right]+(E_g\ \rm independent~terms).
\end{equation}
Using the identity $\log M=\log\det M$ and the fact that $\nu(E_g)$ is a diagonal matrix in the color space thus a simple trace operation, we obtain
\begin{equation}
\log Z=2V\int\frac{d^3\vec{p}}{(2\pi)^2}2\log\prod_a \left[ \sqrt{\L 1-e^{-i\frac{A_a}{T}}e^{\frac{-E_g}{T}}\R\L 1-e^{i\frac{A_a}{T}}e^{\frac{-E_g}{T}}\R} \right],
\end{equation}
where we have ignored the infinite vacuum energy and $E_g$ independent terms. One should notice that each $A_a$ is paired with another $A_b=-A_a$. Eventually, the generating function can be written as a more compacted form:
\begin{equation}\label{weiss}
\log Z=4V{\rm tr_c}\int\frac{d^3\vec{p}}{(2\pi)^2}\log\L 1-\hat{L}_A e^{-E_g/T}\R,
\end{equation}
with $\hat L_A={\rm diag}(\exp(-iA_1/T),\exp(-iA_2/T),...,\exp(-iA_{N^2-1}/T))$ where, again, $\pm A_a$ pairly appear.

\end{document}